\documentclass{aa} 

\usepackage{graphicx}
\usepackage[normalem]{ulem} %Provides Strikeout
\usepackage{txfonts}
\usepackage{amssymb}
\usepackage{enumerate}
\usepackage{subfigure}
\usepackage[colorlinks=true,allcolors=blue]{hyperref}%
\renewcommand{\vec}[1]{{\mathbf #1}}

\newcommand{\avecp}{ \vec A_p}
\newcommand{\avecj}{ \vec A_{\mathrm{j}}}
\newcommand{\bb}{\vec B}

\newcommand{\vv}{ \vec v}

\newcommand{\llnr}[1]{{\bf \color{magenta}{[]}} \color{black}} % to point out where things have been removed by Dupont
\usepackage{amstext}

\begin{document}

 \title{Energy and helicity fluxes in line-tied eruptive simulations}

 \author{L. Linan\inst{1}, \'E. Pariat\inst{1}, G. Aulanier\inst{1}, K. Moraitis\inst{1}, \and G. Valori\inst{2}}

 \institute{LESIA, Observatoire de Paris, Universit\'e PSL, CNRS, Sorbonne Universit\'e, Universit\'e de Paris, 5 place Jules Janssen, 92195 Meudon, France\\
                 \and Mullard Space Science Laboratory, University College London, Holmbury St. Mary, Dorking, Surrey RH5 6NT, UK}

 \date{Received September ; accepted}

% \abstract{}{}{}{}{} 
% 5 {} token are mandatory
 
 \abstract
 {Conservation properties of magnetic helicity and energy in the quasi-ideal and low-$\beta$ solar corona make these two quantities relevant for the study of solar active regions and eruptions.}
 {Based on a decomposition of the magnetic field into potential and nonpotential components, magnetic energy and relative helicity can both also be decomposed into two quantities: potential and free energies, and volume-threading and current-carrying helicities. In this study, we perform a coupled analysis of their behaviors in a set of parametric 3D magnetohydrodynamic (MHD) simulations of solar-like eruptions.}
 {We present the general formulations for the time-varying components of energy and helicity in resistive MHD. We calculated them numerically with a specific gauge, and compared their behaviors in the numerical simulations, which differ from one another by their imposed boundary-driving motions. Thus, we investigated the impact of different active regions surface flows on the development of the energy and helicity-related quantities.}
 {Despite general similarities in their overall behaviors, helicities and energies display different evolutions that cannot be explained in a unique framework. While the energy fluxes are similar in all simulations, the physical mechanisms that govern the evolution of the helicities are markedly distinct from one simulation to another: the evolution of volume-threading helicity can be governed by boundary fluxes or helicity transfer, depending on the simulation.}
 {The eruption takes place for the same value of the ratio of the current-carrying helicity to the total helicity in all simulations. However, our study highlights that this threshold can be reached in different ways, with different helicity-related processes dominating for different photospheric flows. This means that the details of the pre-eruptive dynamics do not influence the eruption-onset helicity-related threshold. Nevertheless, the helicity-flux dynamics may be more or less efficient in changing the time required to reach the onset of the eruption.}


 \maketitle
%
%-------------------------------------------------------------------
\section{Introduction} \label{sec:Introduction}

Magnetic helicity is a volume-integrated ideal magnetohydrodynamic (MHD) invariant describing the level of twist and entanglement of the magnetic field lines. Initially introduced by \citet{Elsasser56}, magnetic helicity is a conserved quantity within the ideal MHD paradigm \citep{Woltjer58}. However, the strict definition of Elsasser is gauge invariant only for magnetically bounded system, a condition that is not satisfied in most cases, such as the solar atmosphere. This led to the introduction by \citet{BergerField84} and \citet{Finn85} of the relative magnetic helicity, a gauge-invariant quantity suitable for use in solar physics and more generally for natural plasmas.

Using a numerical simulation, \citet{Pariat15} confirmed the hypothesis introduced by \citet{Taylor74} that even in presence of nonideal processes, the dissipation of magnetic helicity is negligible. Relative magnetic helicity cannot be dissipated or created within the corona, therefore it can only be transported or annihilated. This conservation property has several major consequences, one of which might be that coronal mass ejections (CMEs) are the consequence of the evacuation of an excess of helicity \citep{Rust94,Low96}.

In recent years, magnetic helicity has been at the heart of many studies dealing with various topics such as the generation of solar eruptions \citep[e.g.,][]{Kusano04,Longcope07b,Priest16}, magnetic reconnection \citep[e.g.,][]{Linton01,Linton02,DelSordo10}, solar filaments \citep[e.g.,][]{Antiochos13,Knizhnik15,ZhaoL15}, and solar and stellar dynamos \citep[e.g.,][]{Brandenburg05,Candelaresi12}.

Magnetic energy is another relevant quantity in MHD with which eruptivity in the solar corona is studied because most of solar events are driven magnetically \citep[e.g.,][]{Schrijver08}. From the point of view of the energetic budget, magnetic energy is the only source of energy that can generate the powerful events that are observed in the solar atmosphere, such as coronal mass ejections, flares, and solar jets \citep{Forbes00}. Magnetic energy can be decomposed into a current-carrying energy, known as free energy, and a potential energy (cf. Sect. \ref{sec:Decompositionenergy}). Solar flares and CMEs are characterized by a rapid change of the coronal magnetic field that does not change the radial component of the photospheric field. Because the potential field is determined by the radial magnetic field at the photosphere, only the free energy can therefore be converted into kinetic and thermal energies during fast coronal events \citep{Aulanier10,Karpen12}. The potential energy thus represents the lowest energy state of the magnetic field in the solar corona \citep[e.g.,][]{Priest14}.

An analysis of magnetic energies combined with a study of magnetic helicities appears a powerful tool for characterizing active regions and their evolutions toward eruptive events. However, measuring these quantities from observational data remains challenging. One possibility is to estimate the accumulation of magnetic helicity and energy in the solar corona by integrating their fluxes across the solar photosphere over time \citep{Kusano02,Nindos03,Yamamoto05,YamamotoSakurai09}. This method cannot trace the coronal evolution and requires high-cadence time-series magnetograms as well as the velocity fields on the photosphere. Because no direct observation of the photospheric velocity is available, it is obtained by inferring the magnetic field on the solar surface. Despite the progress made in deducing the velocity field \citep{Kusano02,Welsch04,Longcope04} as well as further improvement on flux estimations \citep{Pariat05,Chae07,LiuSchuck12,LiuY13,Dalmasse14,Dalmasse18}, the computation of magnetic energy and helicity fluxes remains very sensitive to the method that is used and to the quality of the observations. A different approach is to compute energy and helicity in coronal volumes. Because magnetic energy and helicity are volume integrals, properly computing them with this method requires the full 3D knowledge of the magnetic field in the volume that is studied. Currently, only 2D measurement on the solar surface are provided, therefore a 3D extrapolation of the magnetic field is a necessary step. The diverse methods based on the volume-integration approach for estimating the magnetic relative helicity were benchmarked in \citet{Valori16}. Different solar active regions have previously been investigated \citep{Valori13,Moraitis14,GuoY17,Polito17,Temmer17,James18,Moraitis19,Thalmann19}.

In parallel, the properties of both helicity and energy are still being studied in solar-like parametric simulations. \citet{Berger03} introduced the decomposition of the relative magnetic helicity into two gauge-invariant components: a current-carrying helicity related to the current-carrying magnetic field, and a complementary volume-threading helicity. \citet{Pariat17} followed and estimated these quantities in a set of seven simulations of the formation of solar active regions \citep{Leake13b,Leake14a}. The different simulations led to either stable or eruptive configurations. The authors found that it is possible to distinguish the two configurations by studying the ratio of the current-carrying helicity to the relative helicity. The ratio before the eruption indeed presents high values only in the eruptive case. To better understand the properties of the relative helicity decomposition, \citet{Linan18} provided the first analytical formulae of the time-variation of nonpotential and volume-threading helicity. They also computed and followed them in two simulations of \citet{Leake13b,Leake14a} and in a simulation of the generation of a coronal jet \citep{Pariat05}. They found that the current-carrying helicity does indeed not evolve as a result of boundary fluxes, but builds up through its exchange with the volume-threading helicity. The evolutions of the current-carrying and the volume-threading helicities are partially controlled by a transfer term that reflects the exchange between these two types of helicity. This exchange term dominates the dynamics of the current-carrying helicity at different instants of the simulations. This means that this helicity does not only evolve as a result of boundary fluxes. The eruption phases of these simulation follow the same dynamics: the current-carrying helicity is first transformed into the volume-threading helicity, and then the latter is ejected from the domain by boundary fluxes. Moreover, the transfer term is expressed as a volume integral: consequently, these two helicities are not classically conserved quantities in the sense that they cannot be independently expressed as a flux through the boundaries, even in ideal MHD, unlike the relative magnetic helicity. This finding strengthens the knowledge of the properties of nonpotential and volume-threading helicity that was first studied by \citet{Moraitis14}.

\citet{Zuccarello15} presented 3D parametric resistive MHD simulations of solar coronal eruptions. Simulations are distinguished by the different motions (line-tied) applied on the photosphere with similar but distinct flux cancelation drivers. Their eruptions were driven by the torus instability \citep{Aulanier10,AulanierD10} and occurred at a precise time identified by a series a relaxation runs. Recently, these models were used to investigate the increase in the downward component of the Lorentz force density around an polarity-inversion line in comparison with the photospheric observation \citep{Barczynski19}. From these simulation, \citet{Zuccarello18} studied the impact of the different boundary driving flows on the helicity and energy injection. They found that the helicity ratio of the current-carrying helicity to the relative helicity is clearly associated with the eruption trigger because the eruptions within the different runs took place exactly when the ratio reached the very same threshold value. 

Recently, the first preliminary observational tests confirmed the idea that the helicity ratio is a good predictor of eruptivity. Based on 3D extrapolation and using different nonlinear force-free models, the time evolution of the helicity ratio has been investigated in three active regions: AR~12673 in \citet{Moraitis19}, the most active of the cycle 24; and AR~11158 and AR~12192 in \citet{Thalmann19}, two extensively studied active regions that generated eruptive and confined flares. However, complementary studies are still needed to understand how the different magnetic topologies observed in the solar corona are linked to the dynamics of the helicity ratio.

In the present study, we apply the helicity decomposition to the analysis of the simulations of \citet{Zuccarello15,Zuccarello18} to investigate the time-variations of the different types of magnetic energy and helicity. In particular, we are interested in the way that the different boundary motion influence the helicity and energy dynamics. \citet{Zuccarello18} showed that the different boundary motions lead to different efficiency in injecting helicity and energy in the domain. In the present work, we aim to explain the physical processes that are responsible for these differences: are they related to boundary fluxes, dissipation, or volume evolution? We also examine whether the transfer term between the two helicity components plays a major role in the helicity budgets, as has been observed in \citet{Linan18}. 

Additionally, we study the dynamics of the helicities in comparison with the dynamics of their energy counterparts, for instance, current-carrying helicity and free energy, and volume-threading helicity and potential energy. Our goal is to highlight the differences and the similarities in the helicity and energy buildup. This study aims to improve our knowledge on magnetic helicities and energies, and it is a necessary step to better understand the full topological and energetic complexity of solar active regions.

Our paper is divided into different sections that are organized as follows. First, we present the time-variation of nonpotential and volume-threading helicities (see Sect. \ref{sec:helicity}). In the same way, we then introduce the different components of the magnetic energy and also their time-variation written for the specific case of resistive MHD (see Sect. \ref{sec:energy}). After presenting the simulations (see Sect. \ref{sec:test_cases}), we present the time evolution of the different quantities (see Sect. \ref{sec:Evolution}). Using our set of simulations, we investigate the role of the transfer term between the helicity components in their evolutions (see Sect. \ref{sec:DynHjandHpj}). While we investigate the difference in helicity dynamics in Sect. \ref{sec:descriminating}, we focus on the similarities between magnetic energy and helicity fluxes in Sect. \ref{sec:energyflux}. In the conclusion, we discuss the effect of the different boundary-driving motions on the energy and helicity injection and on the eruptivity helicity ratio.

%--------------------------------------------------------------------
\section{Nonpotential and volume-threading helicities} \label{sec:helicity}

In the fixed volume $V$ bounded by the surface $S$, the magnetic helicity is defined as
\begin{equation} \label{Eq:ClassicH}
H_{m}= \int_{\mathrm{v}}^{} \textbf{A}\cdot\textbf{B} \, \mathrm{d}V,
\end{equation}
with $\textbf{A}$ the vector potential of the studied magnetic field $\textbf{B}$, i.e $\nabla\times\textbf{A}=\textbf{B}$. In practice, this scalar description of the geometrical properties of magnetic field lines is general only if the magnetic field is tangential to the surface, that is, if $V$ is a magnetically bounded volume. The magnetic helicity is gauge invariant if and only if this condition is respected. For the study of natural plasmas, especially in solar physics, the magnetic field does not satisfy this condition, the solar photosphere being subject to significant flux.
\citet{BergerField84} introduced the relative magnetic helicity, a gauge-invariant quantity, based on a reference field. Throughout the paper, we use the potential reference field $\textbf{B}_{\mathrm{p}}$ that has the same normal distribution of $\textbf{B}$ throughout the surface $S$ and satisfies
\begin{equation}
\left\{\begin{array}{l}
\nabla\times\textbf{B}_{\mathrm{p}}=0 \\
\textbf{n}\cdot(\textbf{B}-\textbf{B}_{\mathrm{p}})|_{\mathrm{S}}=0
\end{array} \right.
,\end{equation}
where $\textbf{n}$ is the outward-pointing unit vector normal on $S$. The potential field can thus be defined by a scalar function, such as $\nabla\phi=\textbf{B}_{\mathrm{p}}$, and $\phi$ is the solution of the Laplace equation,
\begin{equation} \label{eq:laplace}
\left\{
\begin{array}{l} \Delta\phi=0 \\
\frac{\partial\phi}{\partial n}|_{\mathrm{S}} = (\textbf{n}\cdot\textbf{B})|_{\mathrm{S}}.
\end{array} \right.
\end{equation}
When $\textbf{A}_{\mathrm{p}}$ is the vector potential of the potential field $\textbf{B}_{\mathrm{p}}=\nabla\times\textbf{A}_{\mathrm{p}}$, the relative magnetic helicity provided by \citet{Finn85} is defined as
\begin{equation}
H_{\mathrm{v}}= \int_{\mathrm{v}}^{} (\textbf{A}+\textbf{A}_{\mathrm{p}})\cdot(\textbf{B}-\textbf{B}_{\mathrm{p}}) \, \mathrm{d}V. \label{eq:h}
\end{equation}
In this form, the relative magnetic helicity is independently invariant to gauge transformation of both $\textbf{A}$ and $\textbf{A}_{\mathrm{p}}$. The difference between the potential field and the magnetic field can be written as a nonpotential magnetic field, $\textbf{B}_{\mathrm{j}}=\textbf{B}-\textbf{B}_{\mathrm{p}}$, associated with the vector $\textbf{A}_{\mathrm{j}}$, defined as $\textbf{A}_{\mathrm{j}}=\textbf{A}-\textbf{A}_{\mathrm{p}}$, such as $\nabla\times\textbf{A}_{\mathrm{j}}=\textbf{B}_{\mathrm{j}}$. When we use this unique decomposition of $\textbf{B}$ and following the work of \citet{Berger03}, Eq. (\ref{eq:h}) can be divided into two gauge-invariant quantities:
\begin{eqnarray} \label{eq:hvdec}
 H_{\mathrm{v}}&=&H_{\mathrm{j}}+H_{\mathrm{pj}}\\
 H_{\mathrm{j}}&=&\int_{\mathrm{v}}^{} \textbf{A}_{\mathrm{j}}\cdot\textbf{B}_{\mathrm{j}} \, \mathrm{d}V \label{eq:hj}\\ 
H_{\mathrm{pj}}&=&2\int_{\mathrm{v}}^{} \textbf{A}_{\mathrm{p}}\cdot\textbf{B}_{\mathrm{j}} \, \mathrm{d}V, \label{eq:hpj}
\end{eqnarray}
where $H_{\mathrm{j}}$ is the current-carrying magnetic helicity associated with only the current-carrying component of the magnetic field $\textbf{B}_{\mathrm{j}}$, and $H_{\mathrm{pj}}$ is the volume-threading helicity involving both $\textbf{B}$ and $\textbf{B}_{\mathrm{p}}$. By construction, both $H_{\mathrm{j}}$ and $H_{\mathrm{pj}}$ are gauge invariant because by virtue of Eq. (\ref{eq:laplace}), $\textbf{B}_{\mathrm{j}}$ has a vanishing normal component on the surface.

In resistive MHD, where $\textbf{E}=-\vv \times \bb +\eta \nabla \times \textbf{B}$ ($\eta$ being the resistivity, which is here assumed to be constant), \citet{Linan18} established the following equation for the time evolution of the current-carrying magnetic helicity $H_{\mathrm{j}}$: 
\begin{eqnarray} \label{eq:dhjdt}
\frac{\text{d}H_{\mathrm{j}}}{\text{d}t}&=&\left.\frac{\text{d}H_{\mathrm{j}}}{\text{d}t}\right|_{\mathrm{Diss}}+\left.\frac{\text{d}H_{\mathrm{j}}}{\text{d}t}\right|_{\mathrm{Bp,\, var}}+\left.\frac{\text{d}H_{\mathrm{j}}}{\text{d}t}\right|_{\mathrm{Transf}} \nonumber\\
&+&F_{\mathrm{Vn,\, Aj}}+F_{\mathrm{Bn,\, Aj}}+F_{\mathrm{Aj,\, Aj}}+F_{\mathrm{\phi,\, Aj}}+F_{\mathrm{Non-ideal,\, Aj}} 
\end{eqnarray}
with
\begin{eqnarray}
 \left.\frac{\text{d}H_{\mathrm{j}}}{\text{d}t}\right|_{\mathrm{Diss}}&=&-2\int_{\mathrm{v}}^{} \eta(\nabla \times \textbf{B})\cdot\textbf{B}_{\mathrm{j}} \, \mathrm{d}V \label{eq:NoId_Aj}\\
 \left.\frac{\text{d}H_{\mathrm{j}}}{\text{d}t}\right|_{\mathrm{Transf}}&=&-2\int_{\mathrm{v}}^{} (\textbf{v}\times\textbf{B})\cdot\textbf{B}_{\mathrm{p}} \, \mathrm{d}V \label{eq:Ftransf_Aj}\\ 
 \left.\frac{\text{d}H_{\mathrm{j}}}{\text{d}t}\right|_{\mathrm{Bp,\, var}}&=&2\int_{\mathrm{v}}^{} \frac{\partial \phi }{\partial t}\nabla\cdot\textbf{A}_{\mathrm{j}} \,\mathrm{d}V \label{eq:Fvar_Aj}\\
 F_{\mathrm{Vn,\, Aj}}&=&-2\int_{\mathrm{S}}^{} (\textbf{B}\cdot\textbf{A}_{\mathrm{j}})\textbf{v}\cdot\mathrm{d}S \label{eq:FVn_Aj}\\
 F_{\mathrm{Bn,\, Aj}}&=&2\int_{\mathrm{S}}^{} (\textbf{v}\cdot\textbf{A}_{\mathrm{j}})\textbf{B}\cdot\mathrm{d}S \label{eq:FBn_Aj}\\
 F_{\mathrm{Aj,\, Aj}}&=&\int_{\mathrm{S}}^{} \textbf{A}_{\mathrm{j}}\times\frac{\partial }{\partial t}\textbf{A}_{\mathrm{j}} \cdot\mathrm{d}S \label{eq:FAj_Aj}\\
 F_{\mathrm{\phi,\, Aj}}&=&-2\int_{\mathrm{S}}^{} \frac{\partial \phi }{\partial t}\textbf{A}_{\mathrm{j}}\cdot \mathrm{d}S \label{eq:Fphi_Aj} \\
 F_{\mathrm{Non-ideal,\, Aj}}&=&-2\int_{\mathrm{S}}^{} \eta(\nabla \times \textbf{B})\times\textbf{A}_{\mathrm{j}}\cdot \mathrm{d}S \label{eq:Fnonideal_Aj}
.\end{eqnarray}
From this decomposition, \citet{Linan18} obtained an equation for the time-variation that is composed only of gauge-invariant terms: 
\begin{eqnarray} \label{eq:dhjdt_gaugeinv}
\frac{\text{d}H_{\mathrm{j}}}{\text{d}t}=\left.\frac{\text{d}H_{\mathrm{j}}}{\text{d}t}\right|_{\mathrm{Own}}+\left.\frac{\text{d}H_{\mathrm{j}}}{\text{d}t}\right|_{\mathrm{Diss}}+\left.\frac{\text{d}H_{\mathrm{j}}}{\text{d}t}\right|_{\mathrm{Transf}},\end{eqnarray}
with
\begin{eqnarray} \label{eq:Own_Hj}
 \left.\frac{\text{d}H_{\mathrm{j}}}{\text{d}t}\right|_{\mathrm{Own}}=&&\left.\frac{\text{d}H_{\mathrm{j}}}{\text{d}t}\right|_{\mathrm{Bp,\, var}}+F_{\mathrm{Non-ideal,\, Aj}} \nonumber \\
 &+&F_{\mathrm{\phi,\, Aj}}+F_{\mathrm{Vn,\, Aj}}+F_{\mathrm{Bn,\, Aj}}+F_{\mathrm{Aj,\, Aj}}.
 \end{eqnarray}
All these terms initially provided by \citet{Linan18} are recalled here because we analyze them and comment on them in the next sections.
 
Similarly, the time evolution of the volume-threading magnetic helicity $H_{\mathrm{pj}}$ can be decomposed as
 \begin{eqnarray} \label{eq:dhpjdt}
\frac{\text{d}H_{\mathrm{pj}}}{\text{d}t}&=&\left.\frac{\text{d}H_{\mathrm{pj}}}{\text{d}t}\right|_{\mathrm{Diss}}+\left.\frac{\text{d}H_{\mathrm{pj}}}{\text{d}t}\right|_{\mathrm{Bp,\, var}} + \left.\frac{\text{d}H_{\mathrm{pj}}}{\text{d}t}\right|_{\mathrm{Transf}} +F_{\mathrm{Non-ideal,\, Ap}} \nonumber\\
&+&F_{\mathrm{Vn,\, Ap}}+F_{\mathrm{Bn,\, Ap}}+F_{\mathrm{Aj,\, Ap}}+F_{\mathrm{\phi,\, Ap}} 
\end{eqnarray}
with
\begin{eqnarray}
 \left.\frac{\text{d}H_{\mathrm{pj}}}{\text{d}t}\right|_{\mathrm{Diss}}&=&-2\int_{\mathrm{v}}^{} \eta(\nabla \times \textbf{B})\cdot\textbf{B}_{\mathrm{p}} \, \mathrm{d}V \label{eq:NoId_Ap}\\
 \left.\frac{\text{d}H_{\mathrm{pj}}}{\text{d}t}\right|_{\mathrm{Transf}}&=&2\int_{\mathrm{v}}^{} (\textbf{v}\times\textbf{B})\cdot\textbf{B}_{\mathrm{p}} \, \mathrm{d}V \label{eq:Ftransf_Ap} \\
 \left.\frac{\text{d}H_{\mathrm{pj}}}{\text{d}t}\right|_{\mathrm{Bp,\, var}}&=&2\int_{\mathrm{v}}^{} \frac{\partial \phi }{\partial t}\nabla\cdot(\avecp-\avecj) \,\mathrm{d}V \label{eq:Fvar_Ap} \\
 F_{\mathrm{Vn,\, Ap}}&=&-2\int_{\mathrm{S}}^{} (\textbf{B}\cdot\textbf{A}_{\mathrm{p}})\textbf{v}\cdot\mathrm{d}S \label{eq:FVn_Ap}\\
 F_{\mathrm{Bn,\, Ap}}&=&2\int_{\mathrm{S}}^{} (\textbf{v}\cdot\textbf{A}_{\mathrm{p}})\textbf{B}\cdot\mathrm{d}S \label{eq:FBn_Ap}\\
 F_{\mathrm{Aj,\, Ap}}&=&2\int_{\mathrm{S}}^{} \textbf{A}_{\mathrm{j}}\times\frac{\partial }{\partial t}\textbf{A}_{\mathrm{p}} \cdot\mathrm{d}S \label{eq:FAj_Ap}\\
 F_{\mathrm{\phi,\, Ap}}&=&-2\int_{\mathrm{S}}^{} \frac{\partial \phi }{\partial t}(\avecp-\avecj)\cdot \mathrm{d}S \label{eq:Fphi_Ap} \\
 F_{\mathrm{Non-ideal,\, Ap}}&=&-2\int_{\mathrm{S}}^{} \eta(\nabla \times \textbf{B})\times\textbf{A}_{\mathrm{p}}\cdot \mathrm{d}S \label{eq:Fnonideal_Ap}
.\end{eqnarray}
The time-variation of $H_{\mathrm{pj}}$ can also be constructed with gauge-invariant terms only:
\begin{eqnarray} \label{eq:dhpjdt_gaugeinv}
\frac{\text{d}H_{\mathrm{pj}}}{\text{d}t}=\left.\frac{\text{d}H_{\mathrm{pj}}}{\text{d}t}\right|_{\mathrm{Own}}+\left.\frac{\text{d}H_{\mathrm{pj}}}{\text{d}t}\right|_{\mathrm{Diss}}-\left.\frac{\text{d}H_{\mathrm{j}}}{\text{d}t}\right|_{\mathrm{Transf}} 
\end{eqnarray}
with
\begin{eqnarray}
\left.\frac{\text{d}H_{\mathrm{pj}}}{\text{d}t}\right|_{\mathrm{Own}}&=&\left.\frac{\text{d}H_{\mathrm{pj}}}{\text{d}t}\right|_{\mathrm{Bp,\, var}}+F_{\mathrm{Non-ideal,\, Ap}} \nonumber\\
&+&F_{\mathrm{Vn,\, Ap}}+F_{\mathrm{Bn,\, Ap}}+F_{\mathrm{Aj,\, Ap}}+F_{\mathrm{\phi,\, Ap}}. \label{eq:Own_Hpj}
\end{eqnarray}
These decompositions were obtained without any particular hypothesis on the gauge that is used. In particular, we are free to use the Coulomb gauge for $\textbf{A}$ and $\textbf{A}_{\mathrm{p}}$. With this choice, the volume terms $\left.\text{d}H_{\mathrm{j}}/\text{d}t\right|_{\mathrm{Bp,\, var}}$ and $\left.\text{d}H_{\mathrm{pj}}/\text{d}t\right|_{\mathrm{Bp,\, var}}$ both vanish. Thus $\left.\text{d}H_{\mathrm{j}}/\text{d}t\right|_{\mathrm{Own}}$ and
$\left.\text{d}H_{\mathrm{pj}}/\text{d}t\right|_{\mathrm{Own}}$ only contain boundary-flux contributions. The transfer term $\left.\text{d}H_{\mathrm{j}}/\text{d}t\right|_{\mathrm{Transf}}$ expresses the exchange between the helicities $H_{\mathrm{j}}$ and $H_{\mathrm{pj}}$ without any consequence on the evolution of the total relative helicity $H_{\mathrm{V}}$. Furthermore, this quantity being a volume term, the time-variations of $H_{\mathrm{j}}$ and $H_{\mathrm{pj}}$ cannot be expressed solely through boundary fluxes. Therefore $H_{\mathrm{j}}$ and $H_{\mathrm{pj}}$ are not conserved quantities in resistive or ideal MHD.

\section{Magnetic energy} \label{sec:energy}
\subsection{Free and potential energies}\label{sec:Decompositionenergy}

With the decomposition of the magnetic field into current-carrying and potential components, $\textbf{B}=\textbf{B}_{\mathrm{p}}+\textbf{B}_{\mathrm{j}}$, the total magnetic energy $\textbf{E}_{\mathrm{v}}$ can be also decomposed as
\begin{eqnarray} \label{eq:energyns}
E_{\mathrm{v}}&=&\frac{1}{8\pi}\int_{\mathrm{v}}^{} \textbf{B}^{2} \, \mathrm{d}V \nonumber\\
&=&E_{\mathrm{p}}+E_{\mathrm{j}}+\frac{1}{4\pi}\int_{S}^{} \phi\textbf{B}_{\mathrm{j}}\cdot \mathrm{d}S-\frac{1}{4\pi}\int_{\mathrm{v}}^{}\phi\nabla\cdot\textbf{B}_{\mathrm{j}}\, \mathrm{d}V,
\end{eqnarray}
with
\begin{equation} 
E_{\mathrm{p}}=\frac{1}{8\pi}\int_{\mathrm{v}}^{} \textbf{B}_{\mathrm{p}}^{2} \, \mathrm{d}V \quad E_{\mathrm{j}}=\frac{1}{8\pi}\int_{\mathrm{v}}^{} \textbf{B}_{\mathrm{j}}^{2} \, \mathrm{d}V. \label{eq:ejep}
\end{equation}
In this decomposition, $E_{\mathrm{j}}$ is the energy of the current-carrying magnetic field, also known as free energy, and $E_{\mathrm{p}}$ the energy of the solenoidal magnetic field. Because the potential field shares the same surface distribution as the total magnetic field, the surface integral vanishes in Eq. (\ref{eq:energyns}). Numerically, the discretization of the mesh grid unavoidably induces a finite level of non-solenoidality ($\nabla\cdot\textbf{B}\ne0$), and consequently, the last term in Eq. (\ref{eq:energyns}) is not exactly null \citep[cf.][]{Valori13}. However, considering a solenoidal field, Eq. (\ref{eq:energyns}) can be simplified into 
\begin{eqnarray} \label{eq:energy}
E_{\mathrm{v}}=E_{\mathrm{j}}+E_{\mathrm{p}}
.\end{eqnarray}
This decomposition is similar to the decomposition of the helicity obtained in Eq. (\ref{eq:hvdec}). However, here, the potential energy $E_{\mathrm{p}}$, unlike the volume-threading helicity, only depends on the potential field without a dependence on the nonpotential field. 
\subsection{Time-variation of the total magnetic energy} \label{time_E}
We aim to determine the time-variation of the total magnetic energy $E_{\mathrm{v}}$ in a fixed volume $V,$ 
\begin{eqnarray}
 \frac{\text{d}E_{\mathrm{v}}}{\text{d}t}&=&\frac{1}{4\pi}\int_{\mathrm{v}}^{} \textbf{B}\cdot\frac{\partial \textbf{B}}{\partial t}\, \mathrm{d}V
.\end{eqnarray}
In the resistive MHD, we use the Faraday law, $\partial \textbf{B}/\partial t = - \nabla \times \textbf{E,}$ and we obtain 
 \begin{eqnarray}
\frac{1}{4\pi}\int_{\mathrm{v}}^{} \textbf{B}\cdot\frac{\partial \textbf{B}}{\partial t}\, \mathrm{d}V &=&-\frac{1}{4\pi}\int_{\mathrm{v}}^{} \textbf{B}\cdot\nabla\times\textbf{E} \, \mathrm{d}V \nonumber \\
 &=&\frac{1}{4\pi}\int_{\mathrm{v}}^{} \textbf{B}\cdot\nabla\times(\textbf{v}\times\textbf{B}) \, \mathrm{d}V \label{eq:dE_dt1}\\
&-&\frac{1}{4\pi}\int_{\mathrm{v}}^{}\textbf{B}\cdot(\nabla\times(\eta\nabla\times\textbf{B})) \, \mathrm{d}V. \nonumber 
\end{eqnarray}
 Using the Gauss divergence theorem, we can decompose the first term of Eq. (\ref{eq:dE_dt1}):
\begin{eqnarray}
\frac{1}{4\pi}\int_{\mathrm{v}}^{} \textbf{B}\cdot\nabla\times(\textbf{v}\times\textbf{B}) \, \mathrm{d}V&=&\frac{1}{4\pi}\int_{S}^{} (\textbf{v}\times\textbf{B})\times\textbf{B} \cdot \mathrm{d}S \\
&+&\frac{1}{4\pi}\int_{\mathrm{v}}^{} (\textbf{v}\times\textbf{B})\cdot(\nabla\times \textbf{B}) \, \mathrm{d}V. \nonumber
\end{eqnarray}
Here, the surface term corresponds to the surface integral of the poynting vector and can be divided into two terms:
 \begin{eqnarray}
\frac{1}{4\pi}\int_{S}^{} (\textbf{v}\times\textbf{B})\times\textbf{B} \cdot \mathrm{d}S=&-&\frac{1}{4\pi}\int_{S}^{} (\textbf{B}\cdot\textbf{B})\textbf{v}\cdot\mathrm{d}S \nonumber \\
&+&\frac{1}{4\pi}\int_{S}^{} (\textbf{v}\cdot\textbf{B})\textbf{B}\cdot\mathrm{d}S.
\end{eqnarray}
Finally, assuming for simplicity that the resistivity is constant in space, the variation of the total magnetic energy can be decomposed as
\begin{eqnarray} \label{eq:dedt}
\frac{\text{d}E_{\mathrm{v}}}{\text{d}t}&=&\left.\frac{\text{d}E_{\mathrm{v}}}{\text{d}t}\right|_{\mathrm{Diss}}+\left.\frac{\text{d}E_{\mathrm{v}}}{\text{d}t}\right|_{\mathrm{var}}+F_{\mathrm{Bn,\, E}}+F_{\mathrm{Vn,\, E}}
,\end{eqnarray}
with
\begin{eqnarray}
 \left.\frac{\text{d}E_{\mathrm{v}}}{\text{d}t}\right|_{\mathrm{Diss}}&=&-\frac{1}{4\pi}\eta\int_{\mathrm{v}}^{}\textbf{B}\cdot(\nabla\times(\nabla\times\textbf{B})) \, \mathrm{d}V \label{eq:FEdiss}\\
\left.\frac{\text{d}E_{\mathrm{v}}}{\text{d}t}\right|_{\mathrm{var}}&=&\frac{1}{4\pi}\int_{\mathrm{v}}^{} (\textbf{v}\times\textbf{B})\cdot(\nabla\times\textbf{B}) \, \mathrm{d}V \label{eq:FEvar}\\
F_{\mathrm{Bn,\, E}}&=&\frac{1}{4\pi}\int_{S}^{} (\textbf{v}\cdot\textbf{B})\textbf{B}\cdot\mathrm{d}S \label{eq:FEbn}\\
F_{\mathrm{Vn,\, E}}&=&-\frac{1}{4\pi}\int_{S}^{} (\textbf{B}\cdot\textbf{B})\textbf{v}\cdot\mathrm{d}S \label{eq:FEvn}
 .\end{eqnarray}
We performed a similar time variation for $E_\mathrm{v}$ as \citet{Linan18} did for helicities as we summarized in Sect. \ref{sec:helicity}. We find two fluxes: $F_{\mathrm{Vn,\, E}}$ a shearing term associated with horizontal motion, and an emerging term $F_{\mathrm{Bn,\, E}}$ that is related to the emergence. \citet{Longcope07a} decomposed the shearing term at the photospheric level into two contributions by differentiating the motion between the different flux patches and the spin motion of isolated flux patches. As with $H_{\mathrm{j}}$ and $H_{\mathrm{pj}}$, the time-variation of $E_{\mathrm{v}}$ cannot be expressed through boundary fluxes, and thus $E_{\mathrm{v}}$ is not a conserved quantity. Even in ideal MHD, when $\left.\text{d}E_{\mathrm{v}}/\text{d}t\right|_{\mathrm{Diss}}$ is null, a volume term $\left.\text{d}E_{\mathrm{v}}/\text{d}t\right|_{\mathrm{var}}$ survives. Unlike $\text{d}H_{\mathrm{j}}/\text{d}t$ and $\text{d}H_{\mathrm{pj}}/\text{d}t$, the time-variation of the total energy $E_{\mathrm{v}}$ depends on neither $\textbf{A}$ nor $\textbf{A}_{\mathrm{p}}$: each term of Eq. (\ref{eq:dedt}) is gauge invariant. 

\subsection{Time-variation of the potential and free magnetic energies}

Similarly to the analysis of $\text{d}E_{\mathrm{v}}/\text{d}t$ in the previous section, it is possible to obtain the time-variation of $E_{\mathrm{p}}$. Using the scalar potential $\phi$ of $\textbf{B}_{\mathrm{p}}$ such as $\nabla\phi=\textbf{B}_{\mathrm{p}}$ and the Gauss divergence theorem, we write
\begin{eqnarray} \label{eq:depdt}
 \frac{\text{d}E_{\mathrm{p}}}{\text{d}t}&=&\frac{1}{4\pi}\int_{\mathrm{v}}^{} \textbf{B}_{\mathrm{p}}\cdot\frac{\partial \textbf{B}_{\mathrm{p}} }{\partial t}\, \mathrm{d}V \nonumber \\
 &=&\frac{1}{4\pi}\int_{\mathrm{v}}^{} \textbf{B}_{\mathrm{p}}\cdot\frac{\partial \nabla\phi }{\partial t}\, \mathrm{d}V \nonumber \\
 &=&F_{\mathrm{\phi, B_{z}}}+\frac{dE_{\mathrm{p}}}{dt}|_{\mathrm{ns}},\end{eqnarray}
 with
\begin{eqnarray}
 F_{\mathrm{\phi, B_{z}}} &=&\frac{1}{4\pi}\int_{S}^{} \frac{\partial \phi }{\partial t}\textbf{B}_{\mathrm{p}}\cdot \mathrm{d}S \label{eq:phiBp} \\
\frac{dE_{\mathrm{p}}}{dt}|_{\mathrm{ns}}&=&-\frac{1}{4\pi}\int_{\mathrm{v}}^{} \frac{\partial\phi }{\partial t}(\nabla\cdot\textbf{B}_{\mathrm{p}})\, \mathrm{d}V. \label{eq:EpNS}
\end{eqnarray}
For a purely solenoidal potential field, $dE_{\mathrm{p}}/dt|_{\mathrm{ns}}$ is null and thus the time-variation of $E_{\mathrm{p}}$ is written in a simple way as a single surface term depending on the variation of the potential field. A decrease in potential energy can therefore be associated in particular with a cancelation at the polarity-inversion line (PIL) or a dispersion of the potential magnetic field. Unlike $E_{\mathrm{v}}$, $E_{\mathrm{p}}$ is a conserved quantity in resistive MHD. 

For the study of the evolution of nonpotential energy $E_{\mathrm{j}}$, the easiest way is to consider only the difference between the decompositions obtained from Eq. (\ref{eq:dedt}) and Eq. (\ref{eq:depdt}):
 \begin{eqnarray} \label{eq:dejdt_ns}
\frac{\text{d}E_{\mathrm{j}}}{\text{d}t}=\frac{\text{d}E}{\text{d}t}-\frac{\text{d}E_{\mathrm{p}}}{\text{d}t}
.\end{eqnarray}
Still, this identity is truly accurate only in the case of purely solenoidal magnetic fields, that is, $\nabla\cdot\textbf{B}_{\mathrm{j}}=0$ and when $\textbf{B}_{\mathrm{j}}|_{S}=0$ (cf Sect. \ref{sec:Decompositionenergy}). Hereafter, we therefore introduce a generic formulation where terms that explicitly account for nonsolenoidal errors are retained,
\begin{eqnarray}
 \frac{\text{d}E_{\mathrm{j}}}{\text{d}t}&=&\frac{1}{4\pi}\int_{\mathrm{v}}^{} \textbf{B}_{\mathrm{j}}\cdot\frac{\partial \textbf{B}_{\mathrm{j}}}{\partial t}\, \mathrm{d}V \\
&=&\frac{\text{d}E_{\mathrm{v}}}{\text{d}t}+ \frac{\text{d}E_{\mathrm{p}}}{\text{d}t}-\frac{1}{4\pi}\int_{\mathrm{v}}^{} \textbf{B}\cdot\frac{\partial \textbf{B}_{\mathrm{p}}}{\partial t}\, \mathrm{d}V-\frac{1}{4\pi}\int_{\mathrm{v}}^{} \textbf{B}_{\mathrm{p}}\cdot\frac{\partial \textbf{B}}{\partial t}\, \mathrm{d}V. \nonumber
\end{eqnarray}
The last two terms on the right are very similar to $dE_{\mathrm{p}}/dt$ and $dE_{\mathrm{v}}/dt$. They can thus be decomposed in the same way,
\begin{equation}
 -\frac{1}{4\pi}\int_{\mathrm{v}}^{} \textbf{B}\cdot\frac{\partial \textbf{B}_{\mathrm{p}}}{\partial t}\, \mathrm{d}V=-
 \frac{1}{4\pi}\int_{S}^{} \frac{\partial \phi }{\partial t}\textbf{B}\cdot \mathrm{d}S+\frac{1}{4\pi}\int_{\mathrm{v}}^{} \frac{\partial\phi }{\partial t}(\nabla\cdot\textbf{B})\, \mathrm{d}V, 
\end{equation}
and
\begin{eqnarray*}
 -\frac{1}{4\pi}\int_{\mathrm{v}}^{} \textbf{B}_{\mathrm{p}}\cdot\frac{\partial \textbf{B}}{\partial t}\, \mathrm{d}V&=&\frac{1}{4\pi}\eta\int_{\mathrm{v}}^{}\textbf{B}_{\mathrm{p}}\cdot(\nabla\times(\nabla\times\textbf{B})) \, \mathrm{d}V \\
&-&\frac{1}{4\pi}\int_{\mathrm{v}}^{} (\textbf{v}\times\textbf{B})\cdot(\nabla\times\textbf{B}_{\mathrm{p}}) \, \mathrm{d}V \\
&-&\frac{1}{4\pi}\int_{S}^{} (\textbf{v}\cdot\textbf{B}_{\mathrm{p}})\textbf{B}\cdot\mathrm{d}S\\
&+&\frac{1}{4\pi}\int_{S}^{} (\textbf{B}\cdot\textbf{B}_{\mathrm{p}})\textbf{v}\cdot\mathrm{d}S.
\end{eqnarray*}
By definition, the curl of the potential field is null, and thus the second volume integral on the right-hand side formally vanishes. Finally, by grouping all the terms, the time-variation of the nonpotential magnetic energy can be written as 
 \begin{eqnarray} \label{eq:dejdt}
\frac{\text{d}E_{\mathrm{j}}}{\text{d}t}&=&\left.\frac{\text{d}E_{\mathrm{j}}}{\text{d}t}\right|_{\mathrm{Diss}}+\left.\frac{\text{d}E_{\mathrm{j}}}{\text{d}t}\right|_{\mathrm{var}}+F_{\mathrm{Bn,\, Ej}}+F_{\mathrm{Vn,\, Ej}} \\
&+&\left.\frac{\text{d}E_{\mathrm{j}}}{\text{d}t}\right|_{\mathrm{ns}}+F_{\mathrm{\phi,\, Bj,}} \nonumber
\end{eqnarray}
with
\begin{eqnarray}
 \left.\frac{\text{d}E_{\mathrm{j}}}{\text{d}t}\right|_{\mathrm{Diss}}&=&-\frac{1}{4\pi}\eta\int_{\mathrm{v}}^{}\textbf{B}_{\mathrm{j}}\cdot(\nabla\times(\nabla\times\textbf{B})) \, \mathrm{d}V \label{eq:FEjdiss}\\
\left.\frac{\text{d}E_{\mathrm{j}}}{\text{d}t}\right|_{\mathrm{var}}&=&\frac{1}{4\pi}\int_{\mathrm{v}}^{} (\textbf{v}\times\textbf{B})\cdot(\nabla\times\textbf{B}) \, \mathrm{d}V \label{eq:FEjvar}\\
F_{\mathrm{Bn,\, Ej}}&=&\frac{1}{4\pi}\int_{S}^{} (\textbf{v}\cdot\textbf{B}_{\mathrm{j}})\textbf{B}\cdot\mathrm{d}S \label{eq:FEjbn}\\
F_{\mathrm{Vn,\, Ej}}&=&-\frac{1}{4\pi}\int_{S}^{} (\textbf{B}\cdot\textbf{B}_{\mathrm{j}})\textbf{v}\cdot\mathrm{d}S \label{eq:FEjvn}\\
F_{\mathrm{\phi,\, Bj}}&=&-\frac{1}{4\pi}\int_{S}^{} \frac{\partial \phi }{\partial t}\textbf{B}_{\mathrm{j}}\cdot \mathrm{d}S \label{eq:FEjphi}\\
\left.\frac{\text{d}E_{\mathrm{j}}}{\text{d}t}\right|_{\mathrm{ns}}&=&\frac{1}{4\pi}\int_{\mathrm{v}}^{} \frac{\partial\phi }{\partial t}(\nabla\cdot\textbf{B}_{\mathrm{j}})\, \mathrm{d}V. \label{eq:EjNs}
\end{eqnarray}
 
As expected, the decomposition we obtain is similar to the one obtained with $dE_{\mathrm{v}}/dt$ (see Eq. \ref{eq:dedt}). The dissipation term $dE_{\mathrm{j}}/dt|_{\mathrm{ns}}$ is null if the magnetic field is solenoidal. However, we compute it in order to quantify the effect of purely numerical errors.

\section{Line-tied eruptive simulations} \label{sec:test_cases}

In order to comparatively analyze the evolution of helicities and energies and to study the time-variation of the energy, we used magnetic field data produced by parametric 3D MHD simulations of eruptive events of the solar corona that were initially presented in \citet{Zuccarello15}.

For this set of simulations, the OHM-MPI code \citep{Aulanier05} solves MHD equations in the system's nondimensional units for a volume covering the domain $x\in [-10,10]$, $y\in [-10,10]$ and $z\in [0,30]$. The employed mesh is nonuniform and the smallest cell is centered at $x=y=z=0$. In order to facilitate the computation of the energies and helicities, our study was performed on a subdomain excluding the $z=0$ plane, interpolated into a uniform Cartesian grid composed of 333 cells in the x and y direction, and 500 cells in the z directions. The resulting analyzed volume is $x\in [-10,10]$, $y\in [-10,10]$ and $z\in [0.006,30]$. As a result of the interpolation on a uniform grid (whose cell sizes are 0.06, to be compared to the original grid, whose cell sizes range from 0.006 to 0.32), the magnetic field we obtained has a lower solenoidality than the initial grid. This reduces the accuracy of the magnetic helicity and energy computations \citep{Valori16}.

The system is delimited by a set of boundaries subject to "open" boundary conditions (except at $z=0$). In the analyzed datasets, the bottom boundary is at $z=0.006$, one mesh point above the surface corresponding to the photospheric level where line-tied boundary conditions were imposed in the original numerical experiments. All the physical MHD quantities, such as the magnetic field, can leave the simulation domain through lateral and top boundaries during the evolution of the system.

Four parametric simulations were performed, all starting with a common phase that is referred to as the "shearing phase" in \citet{Zuccarello15} and \citet{Zuccarello18}. Initially, the magnetic field is potential and generated by two unbalanced and asymmetric subphotospheric polarities. During the shearing phase, asymmetric vortices centered around the local maxima of $|B_{\mathrm{z}}(z=0)|$ slowly evolve the initial potential magnetic field into a current-carrying magnetic field. This shearing flow motion induces a magnetic shear close to the PIL and at the end of this phase, creates a current-carrying magnetic field arcade surrounded by a quasi-potential background field. During this entire phase, the distribution of $B_{\mathrm{z}}$ at the bottom boundary remains unchanged. This phase lasts from the time $t=10t_{\mathrm{A}}$ until $t=100t_{\mathrm{A}}$ in the system time coordinate, where $t_{\mathrm{A}}$ is the reference Alfvén time used in \citet{Zuccarello15}.

\begin{figure*}
 \includegraphics[width=18cm]{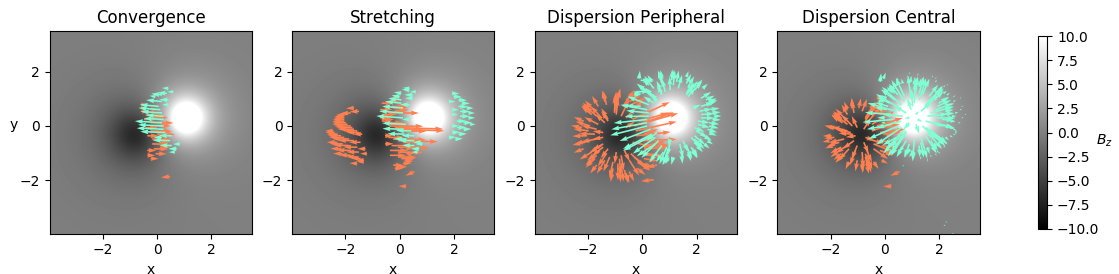}
 \caption{\label{fig:simulation}Applied boundary-driving motions for the four different numerical experiments. White represents the positive polarity ($B_{\mathrm{z}}(z=0.006)>0$) and black the negative polarity ($B_{\mathrm{z}}(z=0.006)<0$). Orange and cyan arrows indicate the distribution of the velocity flows we applied to the negative and positive polarity, respectively.}
\end{figure*}

Then the four parametric simulations differ by the motion pattern that is imposed at the bottom boundary (cf. Fig. \ref{fig:simulation}). We refer to this phase as the "pre-eruption" phase. In a first motion profile, labeled "convergence" (Fig. \ref{fig:simulation}, left panel), the velocity flows only follow the horizontal direction x and are only applied close to the PIL. This creates a cancellation of the magnetic flux around the PIL but only slightly affects the periphery of the active regions. Unlike the previous case, for the run labeled "stretching" (Fig. \ref{fig:simulation}, middle left panel), these horizontal motions are also applied at the periphery of the active region. For the other two runs, labeled "dispersion central" and "dispersion peripheral" (Fig. \ref{fig:simulation}, middle right and right panels), the motions spread in all directions from the center. The only difference is in the portion of the active region that is subjected to these motions. In the dispersion peripheral run, only the periphery of the active region is concerned, while in dispersion central, the dispersion also occurs in the center of the polarity where the magnetic field is strongest.

The four runs all present a cancellation of magnetic flux at the PIL that is permitted by a finite photospheric diffusion. The sheared-arcade configuration at the end of the shearing phase evolves into a bald-patch topology, and the magnetic reconnection process leads to the formation of a flux rope. The system then evolves until it reaches the instant where it becomes unstable and erupts \citep[cf. ][for a description of the eruption process]{Aulanier10}. The onset of the eruption, that is, the time $t_{1}$ of the onset of the instability, is accurately determined by a series of relaxation runs for each simulation \citep{Zuccarello15}. It occurs at $t_{1}$=196, 214, 220 and 164 $t_{\mathrm{A}}$ for the convergence, stretching, dispersion peripheral, and dispersion central runs, respectively. 

To ensure the numerical stability of the code, a finite resistivity $\eta$ and a pseudo-viscosity $\nu$ are necessary \citep{Zuccarello15}. During the common shearing phase until $t<100t_{\mathrm{A}}$ and during the pre-eruption phase, the coronal diffusivity is $\eta_{cor}=4,8.10^{-4}$ and the pseudo-viscosity is fixed to $\nu'=25$. After the eruption, during a phase referred to as eruption phase, $\eta_{cor}$ is $2,1.10^{-3}$ and $\nu'$ is $41,7$. To allow later flux cancellation at the PIL, a photospheric resistivity $\eta_{phot}=\eta_{cor}=4,8.10^{-4}$ is imposed only during the pre-eruption phase. The photospheric resistivity is set to zero in the shearing phase and before the eruption. The change in resistivity follows a ramp-down time profile during the time $t_{1}-5t_{\mathrm{A}}<t<t_{1}+5t_{\mathrm{A}}$. This transitional period is called "eruption onset phase" and is represented as the yellow band in all the figures. We note that the time $t_{1}$ corresponds to the middle of the ramp-down time profile, therefore the boundary flows are null only at $t>t_{1}+5t_{\mathrm{A}}$. In this paper, we removed the first mesh point in the z-direction, which corresponds to the bottom boundary level. We therefore have a uniform resistivity throughout the domain in order to facilitate the calculation of the so-called "nonideal" terms.

\section{Energy and helicity evolutions} \label{sec:Evolution}

In this section we first introduce the method for numerically computing the energies and helicities in our set of simulations. Then, we discuss the computation of the time-variations.

\subsection{Energy and helicity estimations}

In order to compute the different helicities and energies at each time $t_{\mathrm{A}}$, we used the method of \citet{Valori12}. The datacubes of the magnetic field $\textbf{B}$, of the plasma-velocity field $\textbf{v}$, and of the plasma thermodynamic quantities allowed us to compute all the quantities that appear in Eqs. (\ref{eq:dhjdt}), (\ref{eq:dhpjdt}), (\ref{eq:dejdt}), and (\ref{eq:depdt}).

\begin{figure*}
 \subfigure{\label{helicity-1} \includegraphics[width=6.1cm]{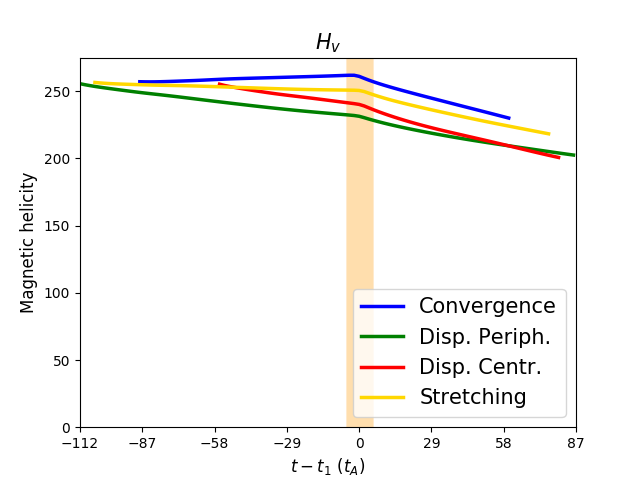}}
 \subfigure{\label{helicity-2} \includegraphics[width=6.1cm]{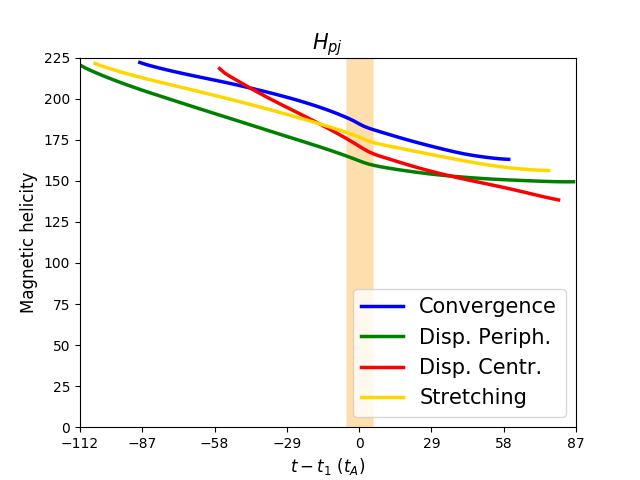}}
 \subfigure{\label{helicity-3} \includegraphics[width=6.1cm]{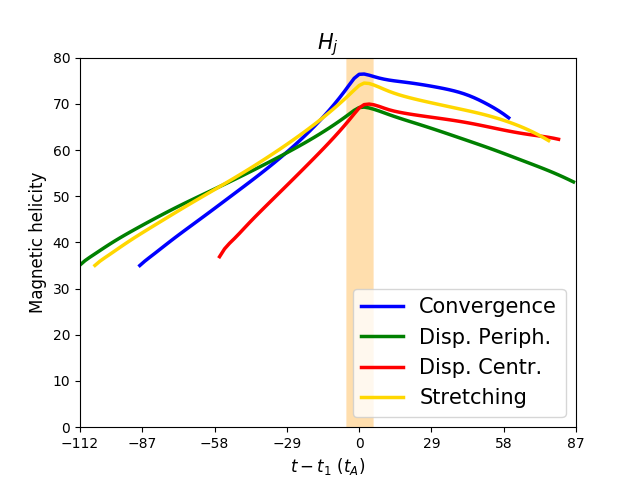}}
 
 \subfigure{\label{energy-4} \includegraphics[width=6.1cm]{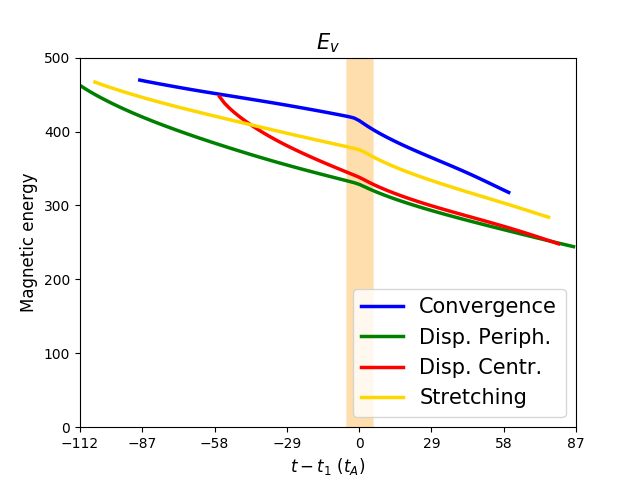}}
 \subfigure{\label{energy-5} \includegraphics[width=6.1cm]{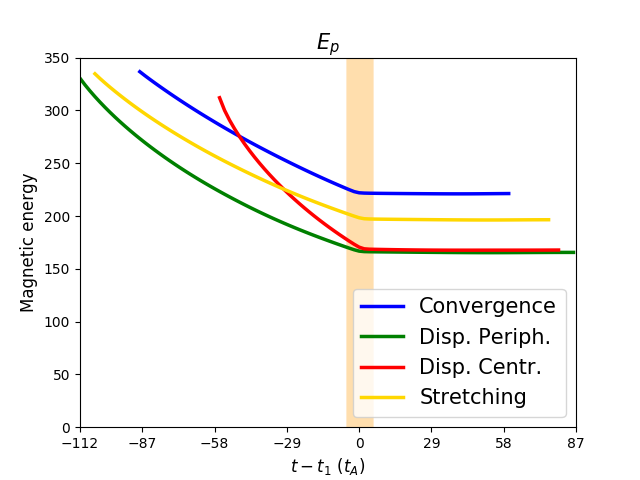}}
 \subfigure{\label{energy-6} \includegraphics[width=6.1cm]{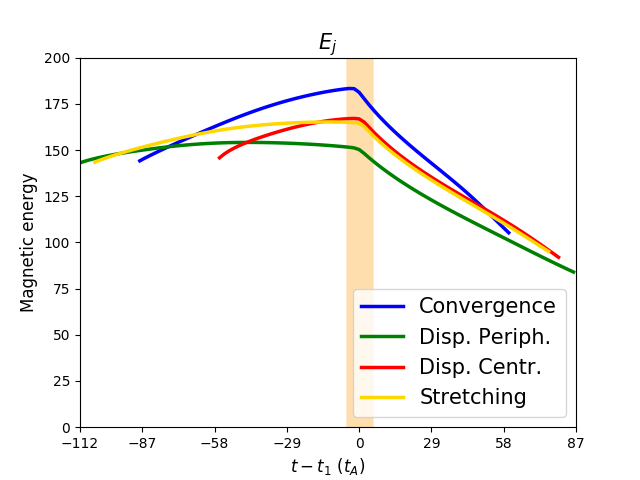}}
 \caption{Evolution of the different magnetic helicities (top panels), from left to right: relative magnetic helicity ( $H_{\mathrm{v}}$, Eq. \ref{eq:h}), volume-threading helicity ($H_{\mathrm{pj}}$, Eq. \ref{eq:hpj}), and current-carrying helicity ($H_{\mathrm{j}}$, Eq. \ref{eq:hj}). Time evolution of the different magnetic energies (bottom panel), from left to right: total magnetic energy ($E_{\mathrm{v}}$, Eq. \ref{eq:energy}), potential energy ($E_{\mathrm{p}}$, Eq. \ref{eq:ejep}), and free energy ($E_{\mathrm{j}}$, Eq. \ref{eq:ejep}). The different simulations are dispersion central (red line), dispersion peripheral (green line), stretching (yellow line) and convergence (blue line). The yellow vertical band corresponds to the onset phase of the eruption. 
 \label{helicity-energy}}
\end{figure*}

First the scalar potential $\phi(t)$ of the potential magnetic field $\textbf{B}_{\mathrm{p}}(t)$ was obtained from a numerical solution of the Laplace equation (\ref{eq:laplace}). The numerical methods we employed to solve this equation required an uniform grid and thus led to the interpolation of the initial grid, as mentioned in Sect. \ref{sec:test_cases}. The potential vectors $\textbf{A}(t)$ and $\textbf{A}_{\mathrm{p}}(t)$ were computed according to Eq. (14) in \citet{Valori12} and follow the DeVore-Coulomb gauge defined in \citet{Pariat15}:
\begin{eqnarray} \label{eq:gauge}
\nabla \cdot \textbf{A}_{\mathrm{p}} &=& 0 \\
A_{\mathrm{z}}(x,y,z,t)&=&A_{\mathrm{p,z}}(x,y,z,t) =0
.\end{eqnarray}
This choice of gauge was complemented by the following relationship inherent in the integration method:
\begin{equation}
A(x,y,z=z_{\mathrm{top}},t)_\perp=A_{\mathrm{p}}(x,y,z=z_{\mathrm{top}},t)_\perp, \label{eq:devore}
\end{equation}
where $\perp$ means the normal component. It corresponds to the 1D integration of magnetic fields starting at the top boundary of the domain at height $z_{\mathrm{top}}$. Finally, we obtained the helicities and energies from Eqs. (\ref{eq:hvdec}) and (\ref{eq:energyns}). In Fig. \ref{helicity-energy} we plot the time evolution of these quantities for the different runs. In order to facilitate the comparison between the different runs, the time is plotted with a modified time variable $t-t_{1}$ in each figure, where $t_{1}$ is the onset time defined in Sect. (\ref{sec:test_cases}) and is different for each of the four simulations.

Unlike \citet{Zuccarello18}, we are interested here in the evolution of the quantities after the common shearing phase, including the eruption phase \citep[which was not studied by][]{Zuccarello18}. We also recall that the domain studied here is slightly different from the one studied in \citet{Zuccarello18} (cf. Sect. \ref{sec:test_cases}).
Fig. \ref{helicity-energy} shows that the dynamics of energies and helicities are qualitatively similar from one simulation to the other during the three different phases. The different boundary-forcing mainly affects the magnitude of the different quantities, but not the quality of their dynamical behaviors.

We also note that the dynamic of helicities and energies changes during the eruption. For the current-carrying helicity, $H_{\mathrm{j}}$, and the free energy, $E_{\mathrm{j}}$, we observe an overall increase during the pre-eruption phase, followed by a decrease during the eruption phase. The quantities related to the potential magnetic fields, $E_{\mathrm{p}}$ and $H_{\mathrm{pj}}$, both decrease in the pre-eruption phase. In the eruption phase $E_{\mathrm{p}}$ remains constant while $H_{\mathrm{pj}}$ continues to decrease, although at a lower rate than in the pre-eruption phase.

Because both $H_{\mathrm{j}}$ and $H_{\mathrm{pj}}$ decrease, the relative helicity $H_{\mathrm{v}}$ also decreases in the eruption phase. The total energy, $E_{\mathrm{v}}$, has a dynamics similar to $H_{\mathrm{v}}$: the system loses energy throughout the simulation, but at a different rate before and after the eruption.

Overall, the behavior of the helicities is similar to their energy counterparts, for example, $H_{\mathrm{v}}$ to $E_{\mathrm{v}}$, $H_{\mathrm{j}}$ to $E_{\mathrm{j}}$, $H_{\mathrm{pj}}$ to $E_{\mathrm{p}}$. This is particularly visible for the current-carrying helicity and the free energy: when $H_{\mathrm{j}}$  increases (decreases), $E_{\mathrm{j}}$ also increases (decreases). In the next sections we focus more on the reasons of these trends by studying the fluxes of the different quantities.

\subsection{Time-variation estimation} \label{sec:validation}

After all the vectors were calculated, we computed the instantaneous time-variations of energies and helicities obtained from Eqs. (\ref{eq:dhjdt}), (\ref{eq:dhpjdt}), (\ref{eq:dejdt}), and (\ref{eq:depdt}). The surface integrals were calculated systematically as the sum of the contributions from the six boundaries. A study focusing on the contribution of the lower boundary alone is conducted in Sect. \ref{sec:Impactflow}.

\citet{Linan18} validated the time-variations equations established for the volume-threading helicity, $H_{\mathrm{pj}}$, and the current-carrying helicity, $H_{\mathrm{j}}$. The accuracy of the computation is related to difference factors such as the discretization and the remapping of the data, spatial and temporal, as well as to nonexplicit numerical diffusive terms that are not accounted for in our analytical resistive MHD model. In our study, we present a complementary test by comparing the time-variation of the relative helicity, $dH_{\mathrm{v}}/dt$ computed from Eq. (23) in \citet{Pariat15}, with the sum of the time-variations of nonpotential and volume-threading helicities, $dH_{\mathrm{j}}/dt+dH_{\mathrm{pj}}/dt$ from (\ref{eq:dhjdt})+(\ref{eq:dhpjdt}). In this way, both sides are computed with the same temporal accuracy.

\begin{figure*}[ht!]
 \centering
 \subfigure{\label{error-2} \includegraphics[width=7.2cm]{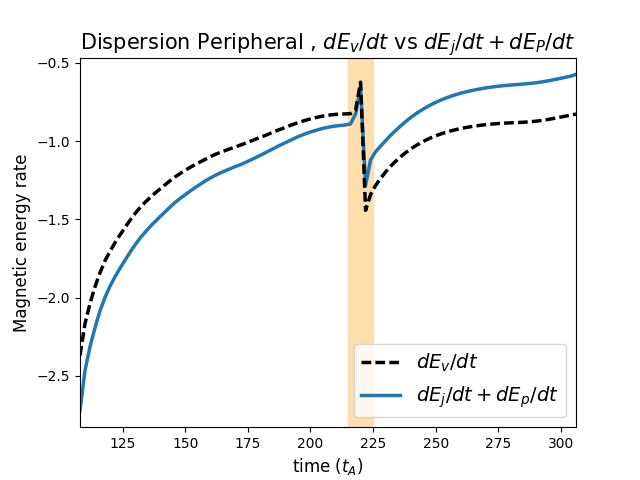}}
 \subfigure{\label{error-3} \includegraphics[width=7.2cm]{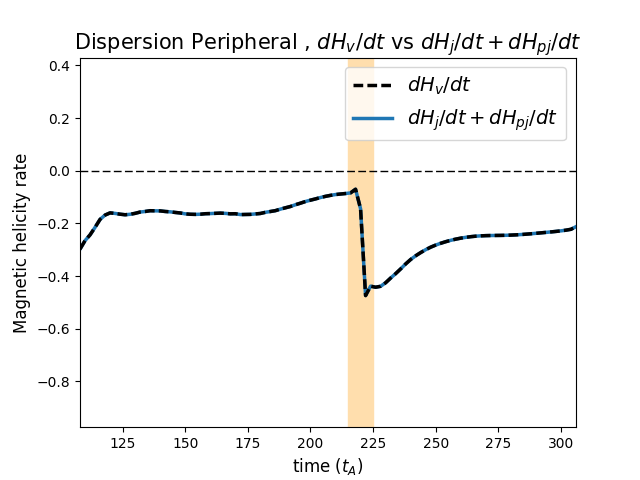}}
 \caption{Left panel: Time evolution of the instantaneous time-variation, $dE_{\mathrm{v}}/dt$ (dashed black curves, Eq. \ref{eq:dedt}), and of the sum, $dE_{\mathrm{j}}/dt+dE_{\mathrm{p}}/dt$ (continuous blue curves, Eqs. (\ref{eq:dhjdt}) and (\ref{eq:dhpjdt})) for the dispersion peripheral run. Right panel: Time evolution of the instantaneous time-variation, $dH_{\mathrm{v}}/dt$ (dashed black curves, Eq. (23) in \citet{Pariat15}, and of the sum, $dH_{\mathrm{j}}/dt+dH_{\mathrm{pj}}/dt$ (continuous blue curves, Eqs. (\ref{eq:dhjdt}) and (\ref{eq:dhpjdt})) for the dispersion peripheral run. The yellow bands correspond to the eruption onset phase.
 \label{error}}
\end{figure*}

The result of this comparison is presented in the right panel of Fig. \ref{error}. In this figure we plot $dH_{\mathrm{v}}/dt$ and $dH_{\mathrm{j}}/dt+dH_{\mathrm{pj}}/dt$ for the dispersion peripheral simulation. For the other runs, the difference is on the same order of magnitude and varies in a similar way. We therefore do not plot this here. The difference is very low, with an average deviation smaller than $0.1\%$. This confirms the robustness of our calculation method as well as the validity of our analytical equations.

In the same way, we compare in Fig. \ref{error} (left panel) the time-variation of the total magnetic energy, $dE_{\mathrm{v}}/dt$, computed from Eq. (\ref{eq:dedt}), with the sum of the time-variations of free and potential energies, $dE_{\mathrm{j}}/dt+dE_{\mathrm{p}}/dt$, from Eqs. (\ref{eq:depdt})+(\ref{eq:dejdt}). Here, the difference is not negligible, with an average relative difference of $27\%$. The cause of this difference is likely mainly the non-solenoidality of the magnetic field. As mentioned Sect. \ref{sec:energy}, with a finite level of non-solenoidality, the equality of Eq. (\ref{eq:dejdt_ns}) is not fully accurate because $E_{\mathrm{v}}\neq E_{\mathrm{j,s}}+E_{\mathrm{p}}$ (see Sect. \ref{sec:Decompositionenergy}).

In order to estimate the artificial non-solenoidal energy contributions, \citet{Valori13} introduced the following decomposition:
\begin{equation}
E_{\mathrm{v}}=E_{\mathrm{j,s}}+E_{\mathrm{p,s}}+E_{\mathrm{p,ns}}+E_{\mathrm{j,ns}}+E_{\mathrm{mix}}
,\end{equation}
where $E_{\mathrm{j,s}}$ and $E_{\mathrm{p,s}}$ are the energies of the current-carrying and potential solenoidal magnetic field. $E_{\mathrm{j,ns}}$ and $E_{\mathrm{p,ns}}$ are the nonsolenoidal components, whereas $E_{\mathrm{mix}}$ corresponds to all the remaining cross terms. For a solenoidal field we have $E_{\mathrm{j,ns}}=E_{\mathrm{p,ns}}=E_{\mathrm{mix}}=0$, $E_{\mathrm{j,ns}}=E_{\mathrm{j}}$, $E_{\mathrm{p,ns}}=E_{\mathrm{p}}$, and therefore $dE_{\mathrm{j}}/dt+dE_{\mathrm{p}}/dt=dE_{\mathrm{v}}/dt$. 

The finite non-solenoidality ($\nabla\cdot\textbf{B}\neq 0)$ affects the precision of the helicity and energy computations, as studied by \citet{Valori16}. Following \citet{Valori13,Valori16,Thalmann19}, we used the energy criteria $E_{\mathrm{div}}/E_{\mathrm{v}}$ to quantify the non-solenoidality effect. The divergence-based energy is defined as\begin{equation}
E_{\mathrm{div}}=E_{\mathrm{p,ns}} +E_{\mathrm{j,ns}}+|E_{\mathrm{mix}}|.
\end{equation}
For the four simulation analyzed in this paper we find an average of $E_{\mathrm{div}}/E_{\mathrm{v}}\simeq2\%$. According to \citet{Valori16}, this value for the average of non-solenoidality leads to a precision of $\leq6\%$ for our helicity computations, which is much lower than the $27\%$ discrepancy found in Fig. \ref{error}. One possible cause of non-solenoidality is our interpolation of the original data from a highly nonuniform mesh onto a uniform grid, which can increase the nondivergence of the magnetic field. Different tests have been made to degrade and also improve the interpolation to give a rough estimate of the effect. Finer interpolations, however, have required considering only a fraction of the whole numerical domain to keep the number of grid points manageable. The outcome of these tests is that neither presented results that differed significantly from our baseline interpolation. In particular, $E_{\mathrm{div}}/E_{\mathrm{v}}$ always remained on the order of $2\%$, which means that this error therefore seems intrinsic to the numerical models. The level of interpolation chosen in our study therefore is a good compromise between the required computed power and the quality of our data. In addition, it is worth noting that various terms in our equations depend on time variations, so that the accuracy of our flux computations can also be limited by a relative coarseness of the time outputs of the available data. Testing for this would require recalculating the simulations with a higher cadence for its outputs, which is beyond the scope of this paper.

\section{Helicity transfer} \label{sec:DynHjandHpj}

\begin{figure*}[ht!]
\centering
 \subfigure{\label{invariantHj-1} \includegraphics[width=8cm]{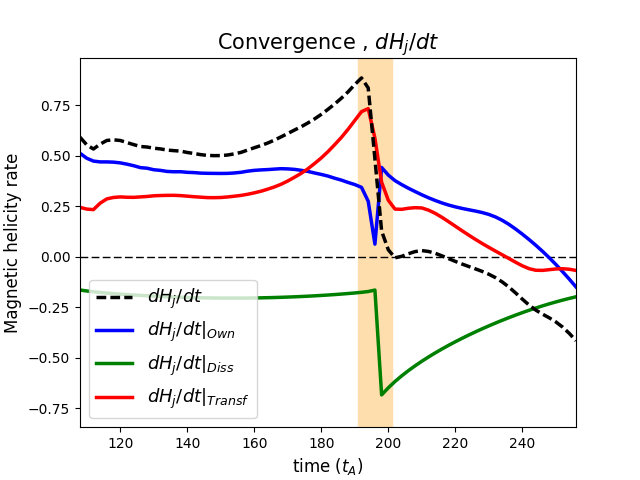}}
 \subfigure{\label{invariantHpj-1} \includegraphics[width=8cm]{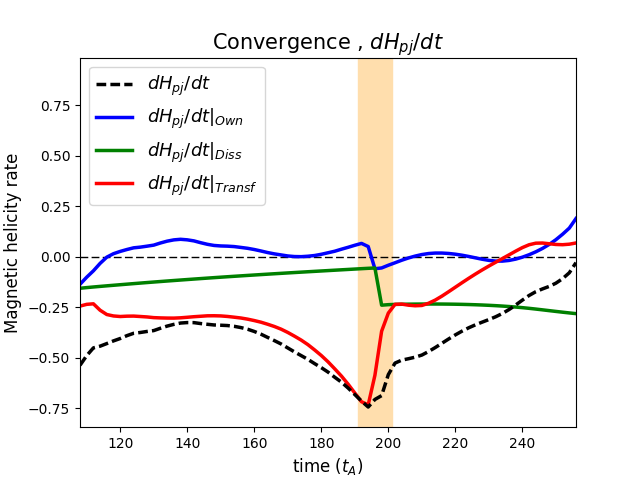}}
 
 \subfigure{\label{invariantHj-2}
 \includegraphics[width=8cm]{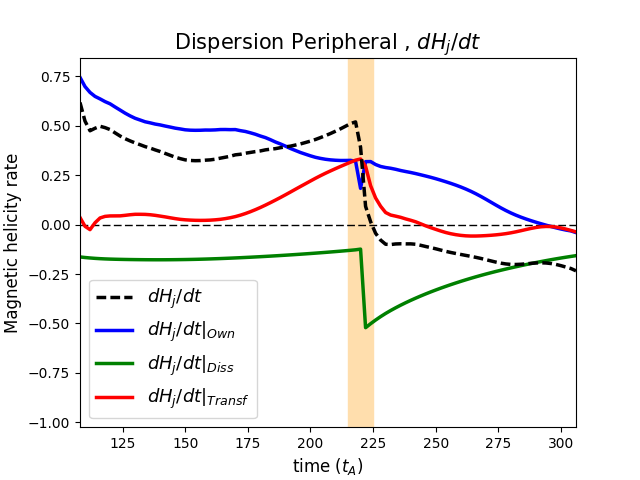}}
 \subfigure{\label{invariantHpj-2} \includegraphics[width=8cm]{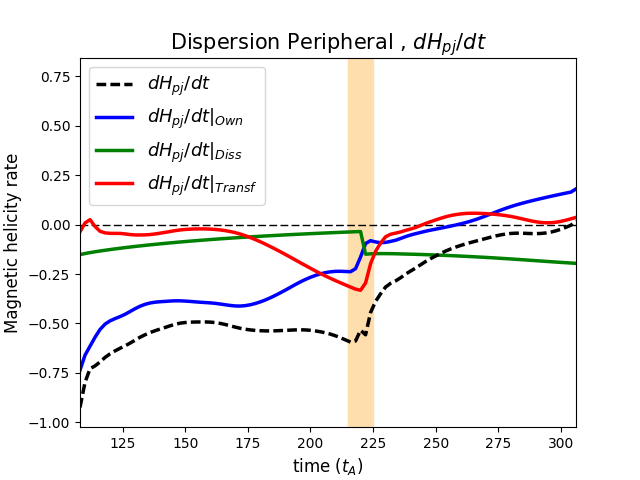}}
 \caption{Time evolution of the helicity variation rates, $dH_{\mathrm{j}}/dt$ and $dH_{\mathrm{pj}}/dt$ (dashed black curves; Eqs. (\ref{eq:dhjdt}) and (\ref{eq:dhpjdt})), of the helicity transfer term, $dH_{\mathrm{j}}/dt|_{\mathrm{Transf}}$ and $dH_{\mathrm{pj}}/dt|_{\mathrm{Transf}}$ (solid red curves; Eqs. (\ref{eq:Ftransf_Aj}) and (\ref{eq:Ftransf_Ap})), of the own terms, $dH_{\mathrm{j}}/dt|_{\mathrm{Own}}$ and $dH_{\mathrm{pj}}/dt|_{\mathrm{Own}}$ (solid blue curves; Eqs. (\ref{eq:Own_Hj}) and (\ref{eq:Own_Hpj})), and of the dissipation terms, $dH_{\mathrm{j}}/dt|_{\mathrm{Diss}}$ and $dH_{\mathrm{pj}}/dt|_{\mathrm{Diss}}$ (solid green curves; Eqs. (\ref{eq:NoId_Aj}) and (\ref{eq:NoId_Ap})) for the convergence simulation (top panels) and for the dispersion peripheral simulation (bottom panels). The left and right panels present the evolution of the current carrying helicity, $H_{\mathrm{j}}$, and volume-threading helicity $H_{\mathrm{pj}}$, respectively.
 \label{invariantH}}
\end{figure*}

As mentioned in the introduction (see Sect. \ref{sec:Introduction}),  \citet{Linan18} showed for two different eruptive simulations that the exchange between $H_{\mathrm{pj}}$ and $H_{\mathrm{j}}$ is controlled by the gauge-invariant transfer term $dH_{\mathrm{j}}/dt|_{\mathrm{Transf}}$ (see Eq. \ref{eq:Ftransf_Aj}), which therefore plays a key role in evolving these helicities. In order to confirm these different results in the particular case of our line-tied simulations, we plot in Fig. \ref{invariantH} the gauge-invariant terms of $dH_{\mathrm{j}}/dt$ (see Eq. \ref{eq:dhjdt_gaugeinv}), and $dH_{\mathrm{pj}}/dt$ (see Eq. \ref{eq:dhpjdt_gaugeinv}) for the convergence and the dispersion peripheral runs.

In the convergence case (see Fig. \ref{invariantH}, top left panel), the conversion of helicity from $H_{\mathrm{pj}}$ to $H_{\mathrm{j}}$ and the boundary flux have similar amplitudes during the pre-eruption phase. Both contributions are positive, thus $H_{\mathrm{j}}$  increases (e.g., $dH_{\mathrm{j}}/dt$ is positive). The transfer term, $dH_{\mathrm{j}}/dt|_{\mathrm{Transf}}$ ,  dominates the evolution of $H_{\mathrm{j}}$ when it is close in time to the eruption. Meanwhile, $H_{\mathrm{pj}}$ decreases mainly because of the strong helicity transfer (cf. Fig. \ref{invariantH}, top right panel), that is, $dH_{\mathrm{j}}/dt|_{\mathrm{Transf}}$ is the dominating term of $dH_{\mathrm{pj}}/dt$. In comparison, the flux of $H_{\mathrm{pj}}$ related to the own term, $dH_{\mathrm{pj}}/dt|_{\mathrm{Own}}$ , is very low. This means that the injection of $H_{\mathrm{pj}}$ is not enough to compensate for its conversion to $H_{\mathrm{j}}$.

Moreover, as with the flux emergence simulations presented in \citet{Linan18}, the eruption is accompanied by a sharp decrease of the transfer term. However, unlike in \citet{Linan18}, the sign of $dH_{\mathrm{j}}/dt|_{\mathrm{Transf}}$ does not change here during the onset phase of the eruption. During the eruption phase, the dissipation terms, $dH_{\mathrm{j}}/dt|_{\mathrm{Diss}}$ and $dH_{\mathrm{pj}}/dt|_{\mathrm{Diss}}$, dominate the variation of the helicities. This is related to the increase in resistivity $\eta$ imposed in the numerical experiment during that period.

In the dispersion peripheral simulation (see Fig. \ref{invariantH}, bottom panels), the variations of $H_{\mathrm{j}}$ and $H_{\mathrm{pj}}$ during the pre-eruption phase are noticeably dominated by the boundary fluxes, $dH_{\mathrm{j}}/dt|_{\mathrm{Own}}$ and $dH_{\mathrm{pj}}/dt|_{\mathrm{Own}}$. The transfer terms are significantly less intense than the injection of $H_{\mathrm{j}}$ and $H_{\mathrm{pj}}$ , except close to the eruption. During the eruption phase, the dynamics is mostly dominated by the resistive dissipation terms, $dH_{\mathrm{j}}/dt|_{\mathrm{Diss}}$ and $dH_{\mathrm{pj}}/dt|_{\mathrm{Diss}}$.

We thus observe that while the trends of $H_{\mathrm{j}}$ and $H_{\mathrm{pj}}$ are similar for the convergence and the dispersion peripheral simulations, as discussed in Sect. \ref{sec:Evolution}, these variations are in fact due to noticeably different dynamics of the helicity fluxes. For instance, the decrease of $H_{\mathrm{pj}}$ in the pre-eruption phase is due to an intense conversion of helicity (high values of $dH_{\mathrm{pj}}/dt|_{\mathrm{Transf}}$) for the dispersion peripheral simulation, whereas in the convergence run, a similar evolution of $H_{\mathrm{pj}}$ is explained by an intense negative boundary flux, $dH_{\mathrm{pj}}/dt|_{\mathrm{Own}}$ during that period.

Finally, as has been noted in \citet{Linan18}, for both the simulations analyzed here but also for the two others, the transfer terms cannot be neglected. We confirm that the estimations of boundary fluxes are not sufficient to follow the dynamics of $H_{\mathrm{j}}$ and $H_{\mathrm{pj}}$. However, unlike with the flux emergence and solar jet simulations studied in \citet{Linan18}, the precise mechanism of the buildup of $H_{\mathrm{j}}$ and $H_{\mathrm{pj}}$ does depend on the simulations during the pre-eruption phase. Studying this dependence is the goal of the next section.

\section{Distinguishing between simulations in terms of helicity dynamics} \label{sec:descriminating}

In the previous section, we described that the magnitude of the different gauge-invariant helicity variation terms could be significantly different in the dispersion peripheral and in the convergence simulations. This demonstrates that even if the general trends of $H_{\mathrm{pj}}$ and $H_{\mathrm{j}}$ are similar (see Sect. \ref{sec:Evolution}), their dynamics can be significantly different.

\begin{figure*}[ht!]
 \subfigure{\label{comparison-1} \includegraphics[width=6.2cm]{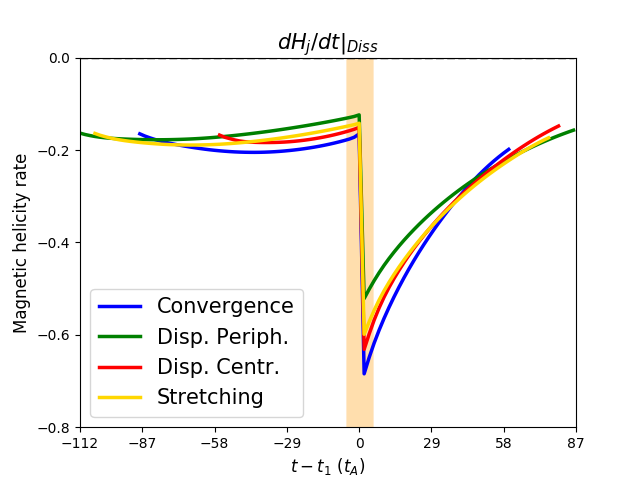}}
 \subfigure{\label{comparison-2} \includegraphics[width=6.2cm]{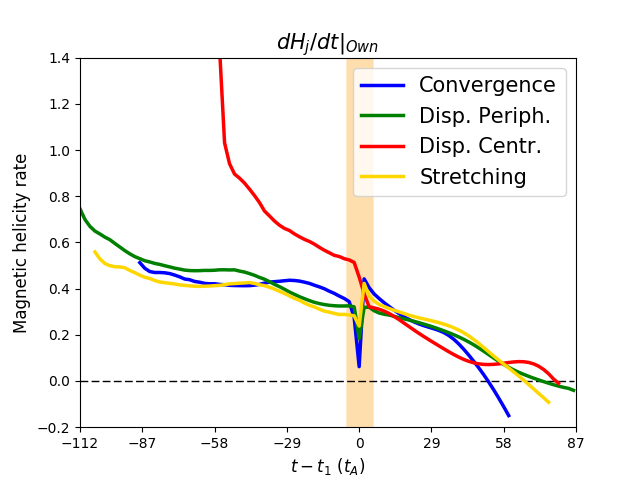}}
 \subfigure{\label{comparison-3} \includegraphics[width=6.2cm]{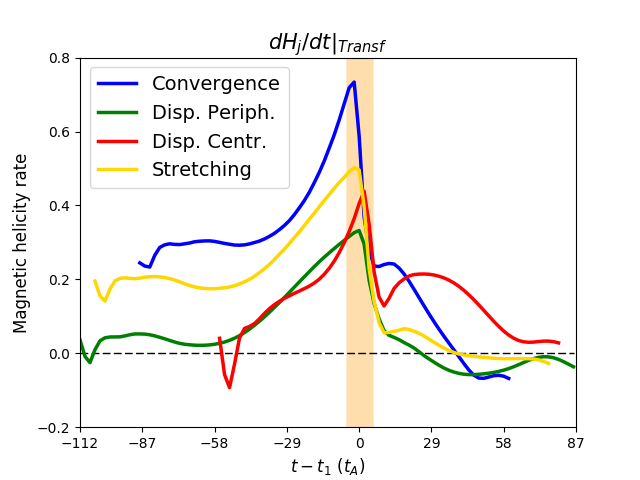}}
 
 \subfigure{\label{comparison-4} \includegraphics[width=6.2cm]{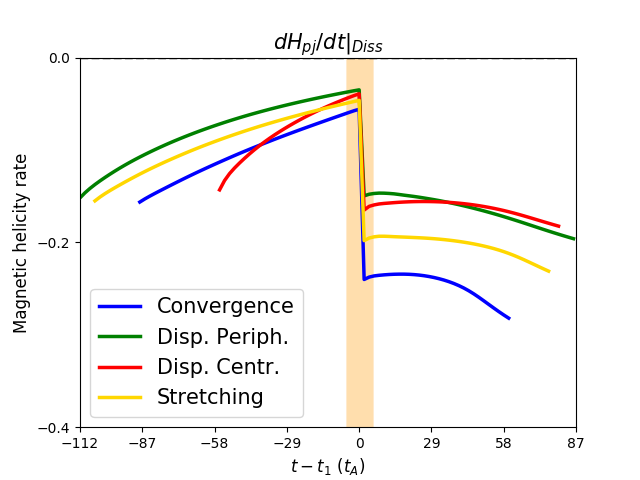}}
 \subfigure{\label{comparison-5} \includegraphics[width=6.2cm]{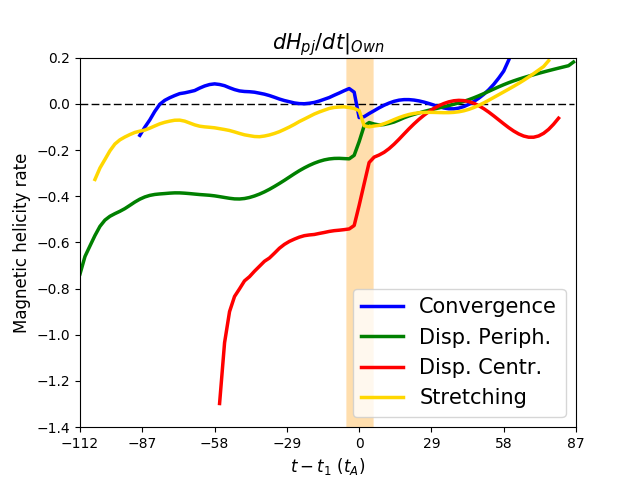}}
 \caption{Time evolution of the different gauge invariant terms of $dH_{\mathrm{j}}/dt$ (top panels), from left to right: dissipation term ($dH_{\mathrm{j}}/dt|_{\mathrm{Diss}}$, Eq. \ref{eq:NoId_Aj}), own term ($dH_{\mathrm{j}}/dt|_{\mathrm{Own}}$, Eq. \ref{eq:Own_Hj}), and helicity transfer term ($dH_{\mathrm{j}}/dt|_{\mathrm{Transf}}$, Eq. \ref{eq:Ftransf_Aj}). Time evolution of the different gauge invariant terms of $dH_{\mathrm{pj}}/dt$ (bottom panels), from left to right: dissipation term ($dH_{\mathrm{pj}}/dt|_{\mathrm{Diss}}$, Eq. \ref{eq:NoId_Ap}), and own term ($dH_{\mathrm{pj}}/dt|_{\mathrm{Own}}$, Eq. \ref{eq:Own_Hpj}). Each curve color corresponds to a particular simulation: dispersion central (red line), dispersion peripheral (green line), stretching (yellow line), and convergence (blue line). The yellow band corresponds to the onset phase of the eruption. 
 \label{fig:comparison}}
\end{figure*}

In order to estimate the effect of the different boundary driver, Fig. \ref{fig:comparison} displays the different gauge-invariant variation terms for the four different numerical simulations: the boundary fluxes $dH_{\mathrm{j}}/dt|_{\mathrm{Own}}$ and $dH_{\mathrm{pj}}/dt|_{\mathrm{Own}}$; the transfer term, $dH_{\mathrm{j}}/dt|_{\mathrm{Transf}}$; and the dissipations terms $dH_{\mathrm{j}}/dt|_{\mathrm{Diss}}$ and $dH_{\mathrm{pj}}/dt|_{\mathrm{Diss}}$.

The dissipations terms (cf. Fig. \ref{fig:comparison}, left panels) are not significantly different from one simulation to the other. The sudden increase in absolute values of the dissipations terms, observed during the eruption onset phase, is related to the imposed numerical increase in resistivity. The variations in magnitude, particularly in the pre-eruption phase, are minor compared to the variations in other gauge-invariant terms.

The boundary flux term $dH_{\mathrm{j}}/dt|_{\mathrm{Own}}$ is also very similar from one simulation to another, except for the dispersion central run, for which more nonpotential helicity is markedly injected during the pre-eruption phase (see Fig. \ref{fig:comparison}, middle top panel). 
Unlike $dH_{\mathrm{j}}/dt|_{\mathrm{Own}}$, the boundary flux of $H_{\mathrm{pj}}$, $dH_{\mathrm{pj}}/dt|_{\mathrm{Own}}$, is strongly sensitive to the boundary-driving pattern (see Fig. \ref{fig:comparison}, bottom left panel). Tthe sign and magnitude of $dH_{\mathrm{pj}}/dt|_{\mathrm{Own}}$ depend on the simulation. For the dispersion simulations, there is a significant injection of negative $H_{\mathrm{pj}}$ , whereas in the convergence and stretching case, the flux is significantly weaker, if not of the opposite sign.

Finally, Fig. \ref{fig:comparison} (top right panel) shows that the helicity transfer rate, $dH_{\mathrm{j}}/dt|_{\mathrm{Transf}}$, is higher for the convergence and stretching simulations than for the dispersion cases in the pre-eruption phase. Unlike with the other cases where the transfer term is almost constant during an early period, in the dispersion central run $dH_{\mathrm{j}}/dt|_{\mathrm{Transf}}$ increases from the first moments of the simulation.

In summary, we observed that during the pre-eruption phase, the increase of $H_{\mathrm{j}}$ and reciprocally the decrease of $H_{\mathrm{pj}}$ (cf Fig. \ref{sec:Evolution}) are not explained by the same physical process in the different simulations. We observe three significantly different dynamics:
\begin{itemize}
\item The convergence and stretching simulations present a similar dynamics for their fluxes of helicity. They are characterized by a relatively weak boundary flux of $H_{\mathrm{pj}}$ counterbalanced by a strong transfer from $H _{\mathrm{pj}}$ to $H_{\mathrm{j}}$. The own term of $H_{\mathrm{j}}$ is positive, involving an injection of current-carrying helicity. Its magnitude is almost identical in these two runs. 
\item For the dispersion peripheral run, $H_{\mathrm{pj}}$ decreases mostly because of the boundary flux, unlike with the previous cases. In comparison to the boundary flux, the transfer from $H_{\mathrm{pj}}$ to $H_{\mathrm{j}}$ is less important. The flux of $H_{\mathrm{j}}$ is similar to the convergence and stretching runs. 
\item The dispersion central shares some similarities with the dispersion peripheral run regarding the variations of $H_{\mathrm{pj}}$. However, this simulation is characterized by a high boundary flux of $H_{\mathrm{j}}$ that is distinct and significantly higher than the three other cases.
\end{itemize}
Finally, simulations with the largest injection of helicities due to their own terms (whether $H_{\mathrm{j}}$ or $H_{\mathrm{pj}}$) have the lowest magnitude of the transfer term. Inversely, a strong exchange between $H_{\mathrm{j}}$ and $H_{\mathrm{pj}}$ is accompanied by lower fluxes through the surfaces. Both lead to a similar trend for the relative helicity $H_{\mathrm{v}}$. This shows that the boundary fluxes of $H_{\mathrm{j}}$ or $H_{\mathrm{pj}}$ as well as the volume term, $dH_{\mathrm{j}}/dt|_{\mathrm{Transf}}$, are directly related to the morphology and the evolution of the magnetic field at the bottom boundary. In Sect. \ref{sec:Impactflow} we discuss that a specific boundary-driven pattern may influence the different physical mechanisms of the evolution of the magnetic helicities. 

\section{Distinguishing between simulations in terms of energy dynamics} \label{sec:energyflux}

As shown in Sect. \ref{sec:Evolution}, the evolutions of $H_{\mathrm{j}}$ and $E_{\mathrm{j}}$ are very similar. Likewise, $H_{\mathrm{pj}}$ and $E_{\mathrm{pj}}$ evolve in the same way during the pre-eruption phase. The main difference appears after the eruption, where $H_{\mathrm{pj}}$ still decreases while $E_{\mathrm{pj}}$ remains constant. However, the similar overall behaviors of $H_{\mathrm{pj}}$ and $H_{\mathrm{j}}$ hide very different physical mechanisms, depending on the simulation, as shown in the previous section. We identified three types of evolution for the dynamics of the helicities. In this section we focus on the following questions: how do $E_{\mathrm{pj}}$ and $E_{\mathrm{j}}$ evolve? Does the dynamics of the energy fluxes also distinguish between the different simulations, as the helicity dynamics do?

\begin{figure*}[ht!]
 \subfigure{\label{fig:fluxEp_C} \includegraphics[width=5.6cm]{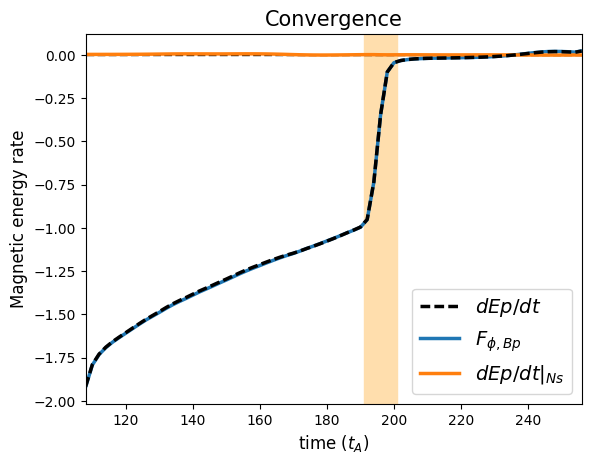}}
 \subfigure{\label{fig:fluxEp_D2} \includegraphics[width=5.4cm]{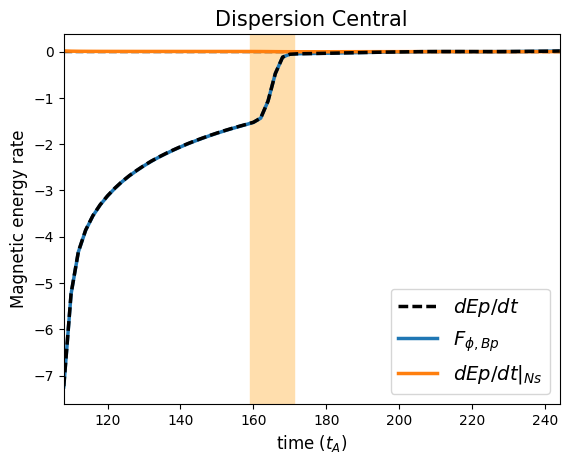}}
 \subfigure{\label{fig:flux_comparisonEp} \includegraphics[width=6.1cm]{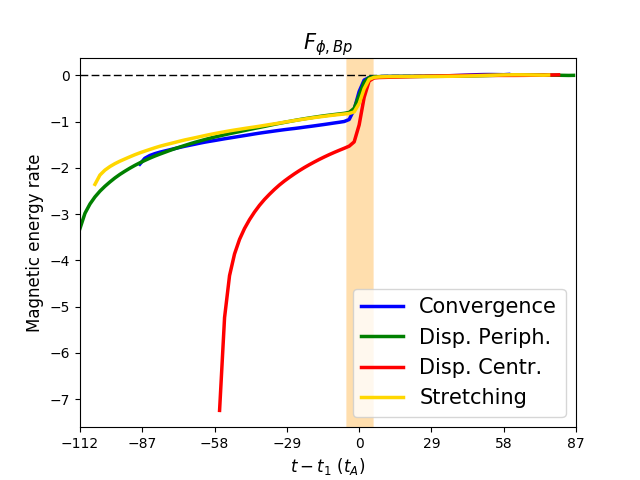}}
 \caption{Left and middle panel: Time evolution of the potential energy variation term (dashed black line; $dE_{\mathrm{p}}/dt$; Equation (\ref{eq:depdt})) and the different terms constituting the instantaneous time-variation of $E_{\mathrm{p}}$ (Eq. \ref{eq:depdt}): $F_{\mathrm{\phi, B_{z}}}$ (blue line; Eq. \ref{eq:phiBp}), and $dE_{\mathrm{p}}/dt|_{\mathrm{ns}}$ (orange line; Eq. \ref{eq:EpNS}). The left and middle panels present the evolution for the convergence and dispersion central simulation, respectively. The right panel presents the time evolution of $F_{\mathrm{\phi, B_{z}}}$ for the four simulations dispersion central (red line), dispersion peripheral (green line), stretching (yellow line), and convergence (blue line). The yellow band corresponds to the onset phase of the eruption. 
 \label{fig:comparisonEp}}
\end{figure*}

For this purpose, we present in Fig. \ref{fig:comparisonEp} all the flux that appear in the decomposition of $dE_{\mathrm{p}}/dt$ (see Eq. \ref{eq:depdt}) for the convergence (Fig. \ref{fig:comparisonEp}, left panel) and the dispersion central simulations (Fig. \ref{fig:comparisonEp}, middle panel). In both simulations, the nonideal term is almost null because of a very low non-solenoidality of the potential field, for instance, $\nabla\cdot B_{\mathrm{p}}\simeq0$. The evolution of the potential energy therefore depends only on $F_{\mathrm{\phi, B_{z}}}$, which results from the evolution of the magnetic field at the boundaries. During the pre-eruption phase, the magnitude of $F_{\mathrm{\phi, B_{z}}}$ as the relative change of the magnetic field at the boundary becomes weaker. Then, during the eruption onset phase, $F_{\mathrm{\phi, B_{z}}}$ decreases strongly before becoming null during the eruption phase. The magnetic field is indeed kept fixed at the bottom boundary during that period.

From comparing $F_{\mathrm{\phi, B_{z}}}$ in the different simulations, we note that $F_{\mathrm{\phi, B_{z}}}$ presents the same evolution for all simulations except for the dispersion central (see Fig. \ref{fig:comparisonEp}, right panel). This indicates that except for the dispersion central simulation, the differences in boundary-driven motions do not affect the injection of the potential energy (cf. the discussion in Sect. \ref{sec:Impactflow}). However, this run has the same functional form as the other, is only more efficient, and therefore quicker, in achieving the eruption.

Regarding $E_{\mathrm{p}}$, only the dispersion central run presents a different behavior. The same conclusion was obtained for the fluxes of $H_{\mathrm{j}}$, for instance, $dH_{\mathrm{j}}/dt|_{\mathrm{Own}}$ (see Fig.\ref{fig:comparison}, middle top panel), but not for $dH_{\mathrm{pj}}/dt|_{\mathrm{Own}}$ (cf. Fig. \ref{fig:comparison}l). First, $F_{\mathrm{\phi, B_{z}}}$ is negative for the entire simulation, while the sign $dH_{\mathrm{pj}}/dt|_{\mathrm{Own}}$ depends on the simulation. This confirms that there is no direct link between the dynamics of $E_{\mathrm{p}}$ and the injection of $H_{\mathrm{pj}}$.

\begin{figure*}[ht!]
 \centering
 \subfigure{\label{FluxEj-1} \includegraphics[width=8cm]{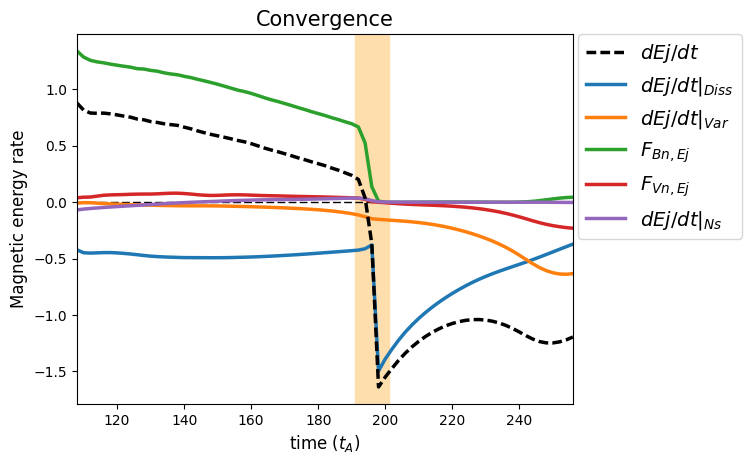}}
 \subfigure{\label{FluxEj-3} \includegraphics[width=8cm]{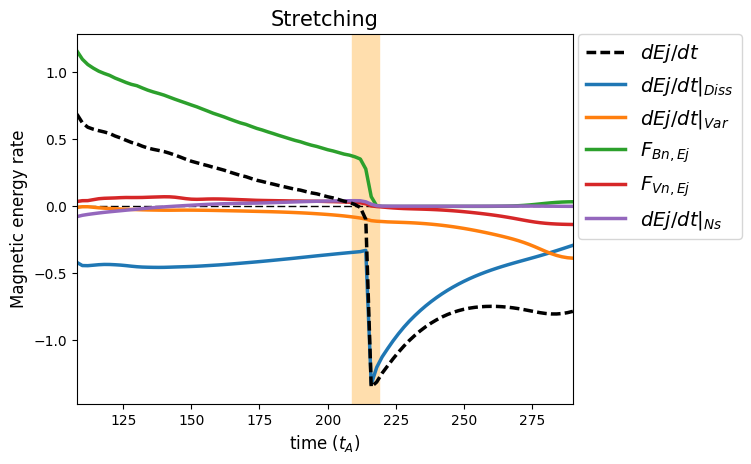}}
 
 \subfigure{\label{FluxEj-4} \includegraphics[width=8cm]{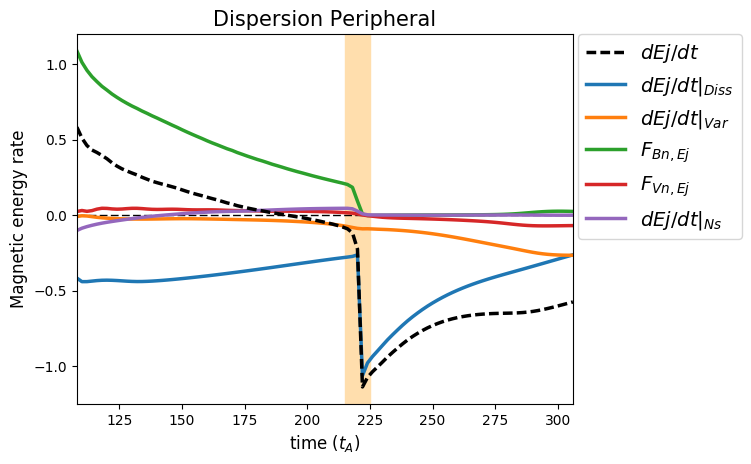}}
 \subfigure{\label{FluxEj-5} \includegraphics[width=8cm]{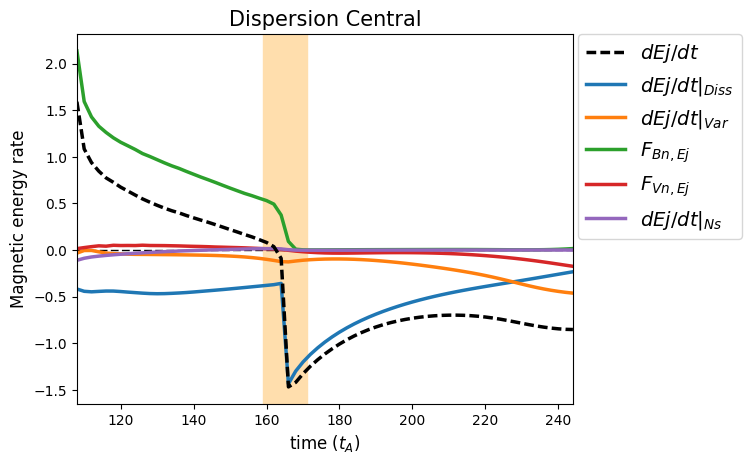}}
 \caption{Time evolution of the free-energy variation rate (dashed black line; $dE_{\mathrm{j}}/dt$; Equation (\ref{eq:dejdt})) and the different terms constituting the instantaneous time-variation of $E_{\mathrm{j}}$ (Eq. \ref{eq:dejdt}): $dE_{\mathrm{j}}/dt|_{\mathrm{Diss}}$ (blue line; Eq. \ref{eq:FEjdiss}), $dE_{\mathrm{j}}/dt|_{Var}$ (orange line; Eq. \ref{eq:FEjvar}), $F_{Bn, Ej}$ (green line; Eq. \ref{eq:FEjbn}), $F_{Vn, Ej}$ (red line; Eq. \ref{eq:FEjvn}), and $dE_{\mathrm{j}}/dt|_{\mathrm{ns}}$ (purple line; Eq. \ref{eq:EjNs}). Each panel corresponds to a different simulation: convergence (top left), stretching (top right), dispersion peripheral (bottom left), and dispersion central (bottom right). The yellow band corresponds to the onset phase of the eruption.
 \label{fig:FluxEj}}
\end{figure*}

In Fig. \ref{fig:FluxEj} we observe the different terms of $dE_{\mathrm{j}}/dt$ (Eq. \ref{eq:dejdt}) for the four simulations. Unlike $dH_{\mathrm{j}}/dt$ and $dH_{\mathrm{pj}}/dt$, the trends and dominant terms of $dE_{\mathrm{j}}/dt$ are similar in the simulations. Only the magnitude of each term may differ. Before the eruption, $dE_{\mathrm{j}}/dt$ is dominated by the emergence term $F_{Bn,E_{\mathrm{j}}}$ despite a significant magnitude of the dissipation term, $dE_{\mathrm{j}}/dt|_{\mathrm{Diss}}$. Then, during the eruption, $F_{Bn,E_{\mathrm{j}}}$ becomes null as a result of the interruption of the boundary-driving flows. During the eruption phase, $dE_{\mathrm{j}}/dt$ is negative and dominated by $dE_{\mathrm{j}}/dt|_{\mathrm{Diss}}$. The free energy mainly decreases because it is dissipated and not ejected out of the volume.

\begin{figure*}[ht!]
 \centering
 \subfigure{\label{comparisonEj-1} \includegraphics[width=8cm]{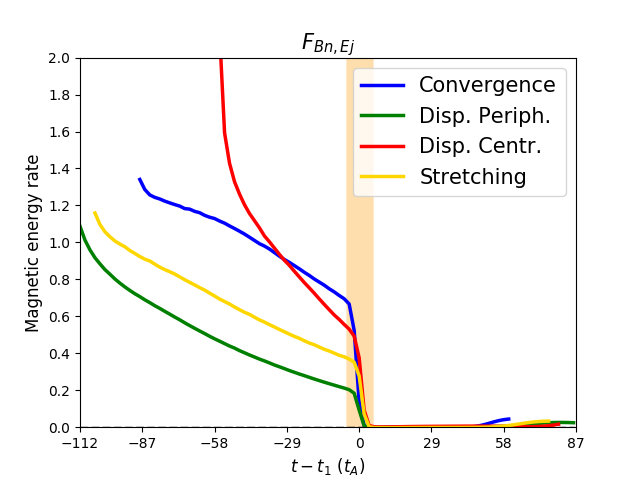}}
 \subfigure{\label{comparisonEj-2} \includegraphics[width=8cm]{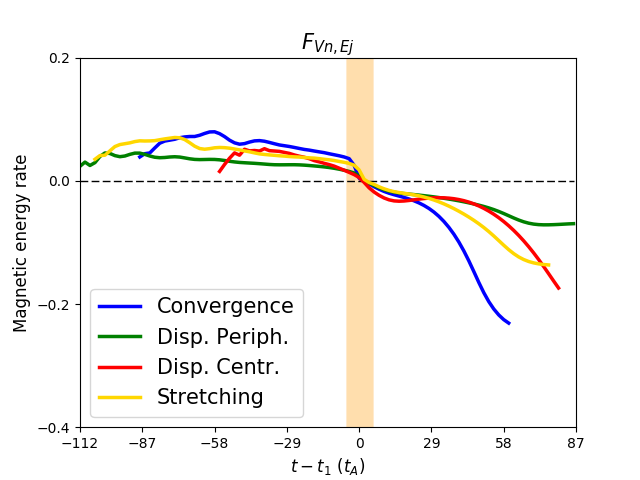}}
 
 \subfigure{\label{comparisonEj-3} \includegraphics[width=8cm]{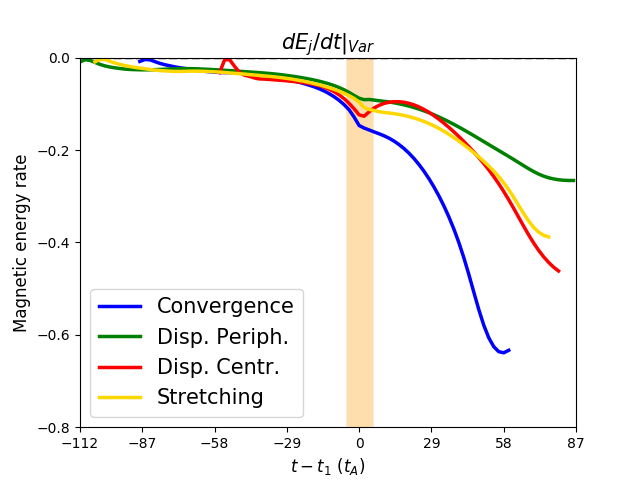}}
 \subfigure{\label{comparisonEj-4} \includegraphics[width=8cm]{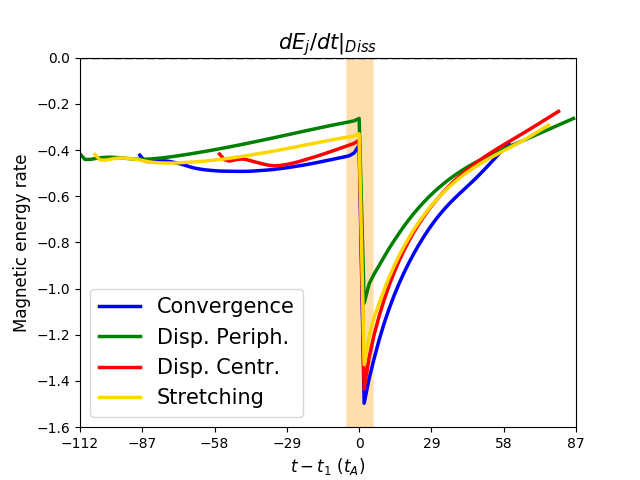}}
 \caption{Time evolution of the different gauge-invariant terms of $dE_{\mathrm{j}}/dt$: $F_{Bn, Ej}$ (top right panel, Eq. \ref{eq:FEjbn}), $F_{Vn, Ej}$ (top left panel, Eq. \ref{eq:FEjvn}), $dE_{\mathrm{j}}/dt|_{Var}$ (bottom left panel, Eq. \ref{eq:FEjvar}), and $dE_{\mathrm{j}}/dt|_{\mathrm{Diss}}$ (bottom right panel, Eq. \ref{eq:FEjdiss}). The different colors present one simulation: dispersion central (red line), dispersion peripheral (green line), stretching (yellow line), and convergence (blue line). The yellow band corresponds to the onset phase of the eruption. 
 \label{fig:comparisonEj}}
\end{figure*}

The dissipation term, $dE_{\mathrm{j}}/dt|_{\mathrm{Diss}}$ does not vary much between the simulations (see Fig. \ref{fig:comparisonEj}, bottom right panel). Similarly, the differences of $F_{Vn,E_{\mathrm{j}}}$ and $dE_{\mathrm{j}}/dt|_{\mathrm{Var}}$ are small between the runs during the pre-eruption phase. Only $F_{Bn,E_{\mathrm{j}}}$ (see Fig. \ref{fig:FluxEj}, top left panel) presents significant differences in the simulations that affect the evolution of the free energy, $E_{\mathrm{j}}$. The dispersion central simulation presents a distinctive trend. The magnitude of $F_{Bn,E_{\mathrm{j}}}$ starts very high and then decreases to values similar to the other runs during the onset phase of the eruption. 

Unlike the evolution of the helicities, $H_{\mathrm{j}}$ and $H_{\mathrm{pj}}$, only the dispersion central simulation stands out from the other runs. This simulation is characterized by a higher decrease of the potential field (see Fig. \ref{fig:comparisonEp}, right panel) and by a higher initial injection of $E_{\mathrm{j}}$ caused by $F_{Bn,E_{\mathrm{j}}}$ (see Fig. \ref{fig:comparisonEj}, top left panel). Before the eruption, another difference with the helicities is that the variations of the trend of $E_{\mathrm{j}}$ and $E_{\mathrm{p}}$ are purely related to the boundary fluxes. Finally, one key outcome of our study is that the dynamics of the energy fluxes do not distinguish between the simulations, unlike the helicity fluxes.

This shows that even if volume helicities and energies follow similar trends (cf. Fig. \ref{helicity-energy}), the physical mechanisms that drive their dynamics  are very different. First, the evolution of free energy, $E_{\mathrm{j}}$ , and potential energy, $E_{\mathrm{p}}$ , are independent, while the current-carrying helicity, $H_{\mathrm{j}}$, evolves in a correlated way with the volume-threading helicity, $H_{\mathrm{pj}}$. Additionally, different boundary forcing only affects the magnitude of the energy fluxes. The dynamics of the helicity is more complex and varies drastically from one simulation to another. One group of simulations (dispersion central and peripheral) is dominated by the flux through the surfaces, while a second group (convergence and stretching) is controlled by volume exchange within the domain. We conclude that energy, helicity, and their decompositions have distinct properties whose analysis should be complementary for the study of the eruptivity of active regions.

\section{Discussion} \label{sec:conclusion}
\subsection{Summary} \label{sec:summary}

In Sect. \ref{sec:helicity} we introduced the formulation of the magnetic relative helicity, $H_{\mathrm{v}}$, as well as the formulation of its decomposition into the current-carrying helicity, $H_{\mathrm{j}}$, and the volume-threading helicity, $H_{\mathrm{pj}}$. We also recalled the analytical equations of their time-variations obtained in \citet{Linan18}. Similarly, we introduced the decomposition of magnetic energy, $E_{\mathrm{v}}$, into the sum of the potential energy, $E_{\mathrm{p}}$, and the free energy, $E_{\mathrm{j}}$. Then, we obtained the time-variation of $E_{\mathrm{v}}$ (see Eq. \ref{eq:dedt}), $E_{\mathrm{p}}$ (see Eq. \ref{eq:depdt}), and $E_{\mathrm{j}}$(see Eq. \ref{eq:dejdt}) by analytically deriving their time derivative (see Sect. \ref{sec:energy}). These formulae are valid for any gauge choices and in the presence of finite level of non-solenoidality for the magnetic field.

Our numerical study of time-variations of energies and helicities is based on a series of four eruptive numerical MHD simulations of solar active regions (see Sect. \ref{sec:test_cases}) that have been investigated in \citet{Zuccarello15}. The evolution of each simulation is characterized by different boundary forcing (line-tied) until the eruption (see Sect. \ref{sec:test_cases}). After the same shearing phase, four driving photospheric flows were considered: convergence, stretching, and peripheral and central dispersion flows. In this study we were particularly interested in the fluxes of energies and helicities during the flux rope formation, during the eruption onset phase, when the torus instability occurs, and during a short time interval after the eruption onset, called the eruption phase. 

Initially, the magnetic energy, $E_{\mathrm{v}}$, decreases as a result of the decrease in the potential magnetic field, $E_{\mathrm{p}}$, despite the increase in free energy, $E_{\mathrm{j}}$ . At the same time, the decrease in volume-threading helicity, $H_{\mathrm{pj}}$, compensates for the injection of the current-carrying helicity, $H_{\mathrm{j}}$, which leads to a quasi-constant evolution of the relative helicity $H_{\mathrm{v}}$ (see Sect. \ref{sec:Evolution}). The relative helicity was mostly injected during the earlier shearing phase.

The fluxes of free and potential energies, $E_{\mathrm{j}}$ and $E_{\mathrm{p}}$ (see Sect. \ref{sec:energyflux}) showed that the effectiveness of the buildup of free energy within the domain is purely related to the magnitude of one surface term. Similarly, the evolution of potential energy is fully linked with its boundary fluxes. 

We also used these simulations to investigate the importance of the exchange between $H_{\mathrm{j}}$ and $H_{\mathrm{pj}}$, which was previously highlighted by \citet{Linan18}. The exchange of helicity between $H_{\mathrm{j}}$ and $H_{\mathrm{pj}}$ is controlled by a gauge-invariant term, $dH_{\mathrm{j}}/dt|_{\mathrm{Transf}}$ (cf. Eq. \ref{eq:Ftransf_Aj}). As in \citet{Linan18}, we observed that this term plays a key role in the dynamics of both $H_{\mathrm{j}}$ and $H_{\mathrm{pj}}$, in particular during the buildup phases where the transfer terms, $dH_{\mathrm{pj}}/dt|_{\mathrm{Transf}}$, dominate the evolution of $dH_{\mathrm{pj}}/dt$ for the runs convergence and stretching. In the dispersion simulations, the evolutions of $dH_{\mathrm{j}}/dt$ and $dH_{\mathrm{pj}}/dt$ are dominated by their boundary fluxes, $dH_{\mathrm{pj}}/dt|_{\mathrm{Own}}$ and $dH_{\mathrm{pj}}/dt|_{\mathrm{Own}}$ (cf. Eqs. \ref{eq:Own_Hj} and \ref{eq:Own_Hpj}), even though the magnitude of the transfer term remains significant. This means that neither $H_{\mathrm{j}}$ nor $H_{\mathrm{pj}}$ evolve as a result of boundary fluxes alone. This conclusion is consistent with the results of \citet{Linan18} that were obtained for different numerical experiments. Specifically, $H_{\mathrm{j}}$ and $H_{\mathrm{pj}}$ cannot be estimated in observed active solar active regions by time integration of its flux through the solar photosphere, but rather with a volume-integration method \citep{Valori16}. This approach requires a 3D reconstruction of the coronal magnetic field from the 2D photospheric measurement with coronal field extrapolation techniques \citep{Wiegelmann12,Wiegelmann14}. A more detailed discussion of the effect of the transfer term on the estimation of $H_{\mathrm{j}}$ and $H_{\mathrm{pj}}$ can be found in the conclusion of \citet{Linan18}.

The key outcome of the study is the observation that the dynamics of the transfer and fluxes of $H_{\mathrm{j}}$ and $H_{\mathrm{pj}}$ depend on the simulation and thus on the imposed driving motions, even though the variations in $H_{\mathrm{j}}$ and $H_{\mathrm{pj}}$ seemed relatively independent of the simulation setup (see Sect. \ref{sec:DynHjandHpj}). Despite the four boundary forcings, the different simulations remain very similar in terms of the magnetic field topology. Nonetheless, the dominant terms of $dH_{\mathrm{j}}/dt$ and $dH_{\mathrm{j}}/dt$ are not the same from one simulation to another.

We highlighted three distinct types of dynamics of the evolution of the helicities in the simulations. In the convergence and stretching simulations, $H_{\mathrm{pj}}$ does not evolve as a result of boundary fluxes, but because of its conversion into $H_{\mathrm{j}}$. The opposite is observed during the two dispersion simulations, for which the evolution of $H_{\mathrm{pj}}$ is mainly related to its fluxes through the boundary, with a weak transfer term. The dispersion central simulation stands out from the others because its boundary flux of the current-carrying helicity, $H_{\mathrm{j}}$, is significantly higher.

Thus we were able to process several photospheric forcings to approach the diversity of active regions that are observed at the solar photosphere. We have come to the conclusion that the evolution of helicity during the formation of a flux rope is a complex process whose origin can be related to fluxes through the surface as well as to volume contributions.

\subsection{Buildup of the helicity ratio} \label{sec:ratio}

\citet{Zuccarello18} have shown that the trigger of the eruptions is related to a threshold in the helicity ratio $H_{\mathrm{j}}/H_{\mathrm{v}}$: this ratio reaches the same value, $|H_{\mathrm{j}}/H_{\mathrm{v}}|_\mathrm{thresh}$ , for all simulations at the onset of the eruptions. In our runs, this threshold is $0.29\pm0.01$. However, as discussed in Sect. 7 of \citet{Zuccarello18}, the exact value of this threshold needs to be taken with care because relative helicity is not simply an additive quantity. We investigated how this helicity ratio is built up and eventually reached by studying the specific dynamics of $H_{\mathrm{j}}$ and $H_{\mathrm{pj}}$ (see Sects. \ref{sec:DynHjandHpj} and \ref{sec:descriminating}). Despite the different boundary forcings, the simulations are very similar, so that it might have been thought that the fluxes of $H_{\mathrm{j}}$ and $H_{\mathrm{pj}}$ would also be similar. However, the key outcome of our study is that the terms that dominate the evolution of $dH_{\mathrm{j}}/dt$ or $dH_{\mathrm{pj}}/dt$ sensitively depend on the simulation even if the overall trends are the same ($H_{\mathrm{j}}$ increases and $H_{\mathrm{pj}}$ decreases, see Sect. \ref{sec:Evolution}). 

Three very distinct ways to reach the helicity eruptivity threshold were found. We observed that the eruption was triggered at a specific value of $H_{\mathrm{j}}/H_{\mathrm{v}}$ independently of the dynamics of $H_{\mathrm{j}}$ and $H_{\mathrm{pj}}$ to reach this threshold. Our work suggests that active regions could reach an eruptive state, either through strong increases of helicity fluxes or through magnetic configurations that induce strong helicity transfer.

The different ways to reach the helicity eruptivity threshold are not all equally effective. The eruption occurs more or less quickly after the end of the shearing phase. The dispersion central run is the most rapid simulation. Then we find the stretching and convergence runs, and last the dispersion peripheral case. The dispersion central simulation stands out from the others because it presents higher helicity fluxes (due to $dH_{\mathrm{j}}/dt|_{\mathrm{Own}}$ and $dH_{\mathrm{pj}}/dt|_{\mathrm{Own}}$) and energy fluxes (due to $F_{\mathrm{\phi, B_{z}}}$ and $F_{Bn,E_{\mathrm{j}}}$) than the other cases.

Finally, using a set of resistive MHD simulations, we provided new  knowledge of the energy and helicity properties. In particular, our analytical and numerical work emphasizes recent studies \citep{Pariat17,Zuccarello18,Moraitis19,Thalmann19} that demonstrated how promising the helicity ratio is as a marker of eruptivity. Further studies are still needed, whether to analytically establish the link between the helicity ratio and torus instability or to properly estimate it from data that are measured in the solar atmosphere.

From direct observational data, the evolution of the ratio $H_{\mathrm{j}}/H_{\mathrm{v}}$ was also analyzed in three active regions with different eruptive profiles. \citet{Moraitis19} investigated the most active region of cycle 24 (AR~12673), while \citet{Thalmann19} focused on an eruptive and a confined flare (AR~11158 and AR~12192). These recent observational analyses seem to qualitatively confirm that the $H_{\mathrm{j}}/H_{\mathrm{v}}$ ratio is tightly related to the eruptivity of solar active regions.
 
\subsection{Effect of the different flows on the helicity and energy injections} \label{sec:Impactflow}

A key result of our study is that the specific driving flows that are applied at the bottom boundary are of significant importance on the dynamics of magnetic helicities and energies. They influence the magnitude and the sign of the own terms for the helicity as well as those of the main fluxes of $dE_{\mathrm{j}}/dt$ and $dE_{\mathrm{p}}/dt$. Moreover, as mentioned in the previous section, even if the way to reach the helicity eruptivity threshold matters less than reaching the threshold, the spatial velocity and magnetic distributions at the boundary affect the time that is required to reach the threshold.

To reach an understanding of the effect of the line-tied forcing, we here briefly discuss the distribution of different quantities at the bottom boundary. A more complete study is beyond the scope of the present paper but will be performed, however. 

\begin{figure*}[ht!]
\centering
\includegraphics[width=17cm]{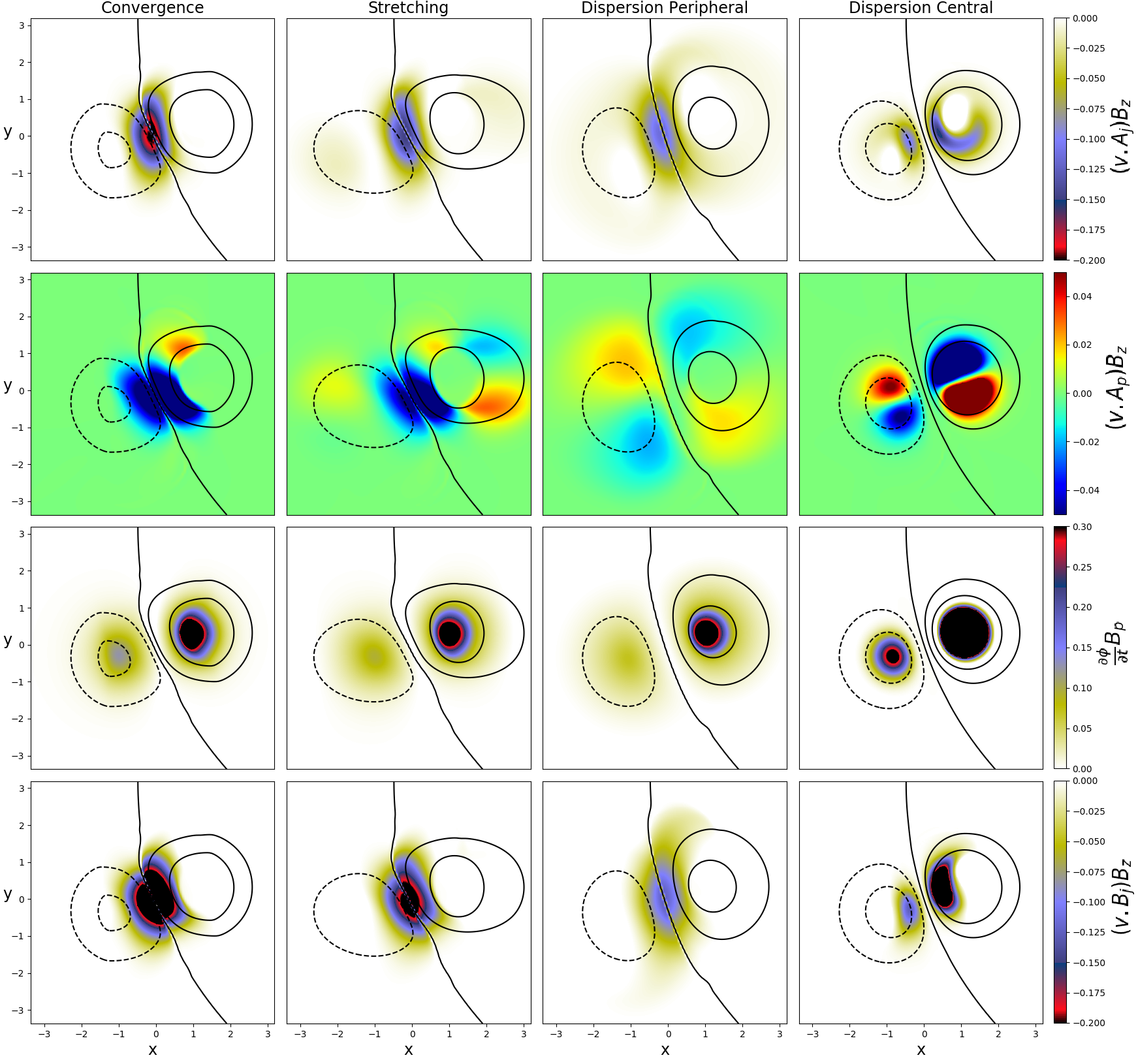}
\caption{From top to bottom, we show the dimensionless magnitude of $(v \cdot A_{\mathrm{j}})B_{\mathrm{z}}$, $(v \cdot A_{\mathrm{p}})B_{\mathrm{z}}$, $(\partial \phi/\partial t)(B_{\mathrm{p}})$, and $(v \cdot B_{\mathrm{j}})B_{\mathrm{z}}$ viewed in the (x, y) plane at $z=0.006$, at the relative time of $t-t_{1}=-58$. Isocontours of $|B_{\mathrm{z}}|$ (dashed line for negative values, solid line for positive values) correspond to values of $|B_{\mathrm{z}}|=-4.5, -2.0, 0, 2.0, \text{and } 4.5$. Each column in the panels presents one simulation, from left to right: convergence, stretching, dispersion peripheral, and dispersion central.}
\label{fig:map}
\end{figure*}

Our goal is to present quantities that might eventually be obtained from observed photospheric magnetograms. Fig. \ref{fig:map} shows four quantities that are related to the energy and helicity fluxes in a (x-y) view at $z=0.006$ for all simulations at the same modified time to the eruption ($t_{1}-t_{\mathrm{A}}=-58$): $(v \cdot A_{\mathrm{j}})B_{\mathrm{z}}$, the integrand of $F_{\mathrm{Vn, Aj}}$, for the injection of $H_{\mathrm{j}}$ (see the first row in Fig. \ref{fig:map}); $(v \cdot A_{\mathrm{j}})B_{\mathrm{z}}$, the integrand of $F_{\mathrm{Vn, Ap}}$, for the injection of $H_{\mathrm{pj}}$ (see the second row in Fig. \ref{fig:map}); $(\partial \phi/\partial t)B_{\mathrm{p}}$, the integrand of $F_{\mathrm{\phi ,Bp}}$ (see the third row in Fig. \ref{fig:map}); and $(v \cdot B_{\mathrm{j}})B_{\mathrm{z}}$, the integrand of $F_{\mathrm{Bn, Ej}}$ (see the fourth row in Fig. \ref{fig:map}). The modified time was taken arbitrarily in the pre-eruption phase.

We recall that the injections of $H_{\mathrm{j}}$ or $H_{\mathrm{pj}}$ related to their own terms cannot be simply reduced to $F_{\mathrm{Vn, Aj}}$ and $F_{\mathrm{Vn, Ap}}$. Other terms such as $dH_{\mathrm{pj}}/dt|_{\mathrm{Transf}}$ must be considered. The second issue is that neither $F_{\mathrm{Vn, Aj}}$ nor $F_{\mathrm{Vn, Ap}}$ are gauge-invariant quantities. They are still good indicators of the spatial distribution of the helicity injection, however, because the same gauge was used for all four runs. Moreover, the emergence term of the relative helicity is one quantity that has traditionally been investigated at the solar photosphere \citep{LiuSchuck12,Bi18}. The distributions of $F_{\mathrm{Vn, Aj}}$ and $F_{\mathrm{Bn, Ej}}$ are very similar (see the first and last row in Fig. \ref{fig:map}). This reveals that the injection of free energy appears to be directly connected to the injection of current-carrying helicity.

The area where $F_{\mathrm{Vn, Aj}}$ and $F_{\mathrm{Bn, Ej}}$ are intense is along the PIL, except for the dispersion central simulation. In the convergence and stretching runs, the shear is favored because the angle between the velocity and the magnetic field at the PIL is high. For the dispersion peripheral case, the velocity is perpendicular to the PIL. Thus the free energy increases more slowly than in the other cases, and more time is needed to build up the flux rope. We also note a low contribution of $F_{\mathrm{Vn, Aj}}$ related to the dispersion of the magnetic field at the periphery of the polarities in the stretching and dispersion peripheral runs.

As highlighted in Sect. \ref{sec:descriminating}, the dynamics in the dispersion central simulation stands out from the others. Unlike the other cases, the intense regions of $F_{\mathrm{Vn, Aj}}$ and $F_{\mathrm{Bn, Ej}}$ are located close to the center of the polarities. The difference between this simulation and the other three appears clearly for $F_{\mathrm{\phi ,Bp}}$ (see the third row in Fig. \ref{fig:map}). In Sect. \ref{sec:energyflux} we observed that $F_{\mathrm{\phi ,Bp}}$ was distinctly higher for the dispersion central than in the other cases. The magnitude of $F_{\mathrm{\phi ,Bp}}$ is higher for the dispersion central simulation because the center of the polarity, where the magnetic field is the most intense, is displaced. The positive polarity, which possess a higher magnetic flux than the negative polarity, contributes more to $F_{\mathrm{\phi ,Bp}}$. The three other cases are very similar because the regions subjected to the flow do not influence $F_{\mathrm{\phi ,Bp}}$.

The distribution of the  $F_{\mathrm{Vn, Ap}}$ term (see the second row in Fig. \ref{fig:map}) is markedly different from the other quantities presented in Fig. \ref{fig:map} . Two different behaviors appear. For the convergence and stretching runs, like $F_{\mathrm{Vn, Aj}}$, the shear along the PIL is the main contribution of $(v \cdot A_{\mathrm{j}})B_{\mathrm{z}}$. For the dispersion runs, $F_{\mathrm{Vn, Ap}}$ is distributed in a symmetric quadrupole at the location where the flow is applied. This leads to an almost null total flux, $F_{\mathrm{Vn, Ap}}$.

Finally, the best way to reach the helicity threshold for these parametric simulations is to facilitate the dispersion of the intense magnetic field. For instance, in the dispersion central case, this corresponds to distributing the flow from the center of each polarity. For the other cases with lower dispersion, the efficient ways to build up the energies and helicities are mostly linked with flux-cancellation converging motions that contain a shearing-flow component parallel to the PIL. The convergence and stretching runs have more intense contributions there.

The observations made above offer only a limited insight on the effect of the different flows on the evolution of energy and helicity. We saw that even if the four simulations look similar overall, some differences related to the applied motions clearly appear in the dynamic of the helicity and energy. New investigations are required to provide a better understanding of the energy and helicity dynamics. Our study especially provides new information for interpreting the injection of helicity and energy in observed active regions. 

With HMI vector magnetic field data, the evolution of energy and helicity flux has previously been studied in different active regions \citep{LiuSchuck12}. This is commonly based on the decomposition of the flux into two components: a shear component provided by photospheric tangential flow, and a vertical component linked with normal flows due to emergence \citep{Bi18}. At the same time, new methods for properly measuring helicities in the solar corona  are still being developed \citep{Dalmasse13,Dalmasse14,Dalmasse18,Valori16,GuoY17,Moraitis18,Gosain19}. For instance, new analytic expressions for the helicity transport allow us to estimate the injection of relative magnetic helicity into the solar atmosphere over an entire solar cycle \citep{Hawkes19,Pipin19}. However, the quality of the helicity and energy estimations from observational data greatly depends on the accuracy of the magnetic field measurements. In particular, it is still difficult to measure the tangential component of the magnetic field. The instrumentation on board the new Solar Orbiter mission, for example, the PHI magnetogram, will help us to make significant progress on the magnetic field measurement and consequently on the helicity and energy computations. Future results that will benefit from our study will provide new insight for a better understanding of the solar coronal activity.

\begin{acknowledgements}
The authors thank the anonymous reviewer for careful reading of the manuscript and helpful comments. L. L., E. P., K. M. acknowledge support of the French Agence Nationale pour la Recherche through the HELISOL project ANR-15-CE31-0001. L. L., E. P acknowledge the support of the french Programme National Soleil-Terre. GV acknowledges the support  from the European Union's Horizon 2020 research and innovation programme under grant agreement No 824135 and of the STFC grant number ST/T000317/1. The numerical simulation used in this work was  performed on the HPC center MesoPSL financed by the project Equip@Meso (reference ANR-10-EQPX-29-01) and the Region Ile-de-France. This article profited from discussions during the meetings of the ISSI International Team Magnetic Helicity in Astrophysical Plasmas. 
\end{acknowledgements}

% \bibliographystyle{aa} % style aa.bst
% \bibliography{bibliography.bib} % your references Yourfile.bib

%\begin{thebibliography}{}

\bibliographystyle{aa} % style aa.bst
\bibliography{biblio.bib}

%\end{thebibliography}{}

%%%%%%%%% FIG 1

%%%%%%%%% FIG 2

%%%%%%%%% FIG 3

%%%%%%%%% FIG 4
 
%%%%%%%%% FIG 5

%%%%%%%%% FIG 6

%%%%%%%%% FIG 7

%%%%%%%%% FIG 8
%%%%%%%%% FIG 9

\end{document}